\documentclass[paper=A4,11pt,DIV=12,headings=big]{scrartcl}
\pdfoutput=1

\usepackage[a4paper, top=1.2in, bottom=1.2in, left=1.1in, right=1.1in]{geometry}
\usepackage{amsfonts} 
\usepackage{amsmath,amssymb}
\usepackage{mathrsfs}
\usepackage[normal]{caption}
\usepackage{slashed}

\usepackage{cancel}
\usepackage{xcolor}
\usepackage{xspace}
\usepackage{cite}
\usepackage{enumerate}
\usepackage{graphicx}
\graphicspath{{./Figs/}}
\usepackage{wrapfig}

\renewenvironment{abstract}{%
\begin{minipage}{0.95\textwidth}
}
{\par\noindent\end{minipage}}

\usepackage[multiple]{footmisc}
\let\oldfootnote\footnote\renewcommand\footnote[1]{\oldfootnote{\hspace{2mm}#1}}

\definecolor{darkblue}{rgb}{0,0,0.9}
\usepackage{hyperref}
\hypersetup{
linktocpage,
colorlinks,
citecolor=darkblue,
filecolor=darkblue,
linkcolor=darkblue,
urlcolor=darkblue
}

\allowdisplaybreaks
\addtokomafont{disposition}{\rmfamily\boldmath}

\newcommand{\mc}{\mathcal}

\newcommand{\<}{\langle}
\renewcommand{\>}{\rangle}

\newcommand{\sh}{q^2 / M_{B_s}^2}

\newcommand{\matel}[3]{\langle #1|#2|#3\rangle}
\newcommand{\al}{\alpha}
\newcommand{\be}{\beta}
\newcommand{\ga}{\gamma}
\newcommand{\de}{\delta}
\newcommand{\la}{\lambda}
\newcommand{\eps}{\epsilon}
\newcommand{\Pperp}[1]{P_\perp^{#1}}
\newcommand{\Ppara}[1]{P_\parallel^{#1}}
\newcommand{\FV}{V_\perp}
\newcommand{\FA}{V_\parallel}
\newcommand{\FVA}{V_{\perp,\parallel}}

\newcommand{\FTV}{T_\perp}
\newcommand{\FTVb}{\overline{T}_\perp}
\newcommand{\FTA}{T_\parallel}
\newcommand{\FTAb}{\overline{T}_\parallel}
\newcommand{\FTVA}{T_{\perp,\parallel}}
\newcommand{\FTVAb}{\overline{T}_{\perp,\parallel}}

\newcommand{\APP}{\mathfrak{a}}
\newcommand{\APPb}{\APP_{\perp,\parallel}}
\newcommand{\APPperp}{\APP_\perp}
\newcommand{\APPpara}{\APP_\parallel}

\def\mumu{\ensuremath{\mu^{+} \mu^{-}}\xspace}
\def\bsmumugamma{\ensuremath{B^0_s \to \mu^{+} \mu^{-} \gamma}\xspace}
\def\bsllgamma{\ensuremath{B^0_s \to \ell^{+} \ell^{-} \gamma}\xspace}

\def\bseegamma{\ensuremath{B^0_s \to e^{+} e^{-} \gamma}\xspace}

\newcommand{\RKst}{\ensuremath R_{K^{0*}}\xspace}

\def\sla#1{\setbox0=\hbox{$#1$}\dimen0=\wd0
      \setbox1=\hbox{/} \dimen1=\wd1 \ifdim\dimen0>\dimen1
      \rlap{\hbox to \dimen0{\hfil/\hfil}} #1                        \else
      \rlap{\hbox to \dimen1{\hfil$#1$\hfil}}
      /   \fi}

\newcommand{\beq}{\begin{equation}}
\newcommand{\eeq}{\end{equation}}
\newcommand{\beqa}{\begin{eqnarray}}
\newcommand{\eeqa}{\end{eqnarray}}

\newcommand{\nn}{\nonumber}

\DeclareOldFontCommand{\rm}{\normalfont\rmfamily}{\mathrm}
\DeclareOldFontCommand{\sf}{\normalfont\sffamily}{\mathsf}
\DeclareOldFontCommand{\tt}{\normalfont\ttfamily}{\mathtt}
\DeclareOldFontCommand{\bf}{\normalfont\bfseries}{\mathbf}
\DeclareOldFontCommand{\it}{\normalfont\itshape}{\mathit}
\DeclareOldFontCommand{\sl}{\normalfont\slshape}{\@nomath\sl}
\DeclareOldFontCommand{\sc}{\normalfont\scshape}{\@nomath\sc}

\begin{document}


\begin{flushright}
LAPTH-025/17 \\
ZU-TH 23/17 \\
CP3-Origins-2017-030 DNRF90
\end{flushright}

\vskip1.0cm

\begin{center}
{\sffamily \bfseries \LARGE \boldmath $B^0_s \to \ell^+ \ell^- \gamma$ as a Test of Lepton Flavor Universality}\\[0.8 cm]
{\normalsize \sffamily \bfseries Diego Guadagnoli$^a$, M\'eril Reboud$^{a,b}$ and Roman Zwicky$^{c,d}$} \\[0.5 cm]
\small
$^a${\em Laboratoire d'Annecy-le-Vieux de Physique Th\'eorique UMR5108\,, Universit\'e de Savoie Mont-Blanc et CNRS, B.P.~110, F-74941, Annecy-le-Vieux Cedex, France}\\[0.1cm]
$^b${\em \'Ecole Normale Sup\'erieure de Lyon, F-69364, Lyon Cedex 07, France}\\[0.1cm]
$^c$ {\em School of Physics and Astronomy, University of Edinburgh, 
    Edinburgh EH9 3JZ, Scotland}\\[0.1cm]
$^d${\em Department of Physics, Universit\"at Zu\"rich, Winterthurerstrasse 190, CH-8057 Z\"urich, Switzerland}\\[0.1cm]
\end{center}

\medskip

\begin{abstract}\noindent

We discuss a number of strategies to reduce the $\mathcal B(B^0_s \to \ell^{+} \ell^{-} \gamma)$ theoretical error, and make such a measurement a new probe of the interactions that are interesting in the light of present-day flavor discrepancies. In particular, for low di-lepton invariant mass we propose to exploit the close parenthood between $\mathcal B(B^0_s \to \ell^{+} \ell^{-} \gamma)$ and the measured $\mathcal B(B^0_s \to \phi (\to K^+ K^-) \gamma)$. For high $q^2$, conversely, we exploit the fact that the decay is dominated by two form-factor combinations, plus contributions from broad charmonium that we model accordingly. We construct the ratio $R_\gamma$, akin to $R_K$ and likewise sensitive to lepton-universality violation. Provided the two rates in this ratio are integrated in a suitable region that minimises bremsstrahlung contributions while maximising statistics, the ratio is very close to unity and the form-factor dependence cancels to an extent that makes it a new valuable probe of lepton-universality violating contributions in the effective Hamiltonian. We finally speculate on additional ideas to extract short-distance information from resonance regions, which are theoretically interesting but statistically limited at present.

\end{abstract}

\vspace{1.2cm}

\renewcommand{\thefootnote}{\arabic{footnote}}
\setcounter{footnote}{0}

\tableofcontents

\newpage

\section{Introduction} \label{sec:intro}

\noindent Flavor data from different experiments display persistent anomalies in $b \to s$ and $b \to c$ decays, obeying a consistent pattern. The very first feature to be remarked is qualitative, and is the fact that a whole range of differential branching fractions of $b \to s \mumu$ modes measured at LHCb lie below the respective Standard-Model (SM) prediction. This is the case for the following channels: $B^{0} \to K^{0} \mumu$, $B^{+} \to K^{+} \mumu$, $B^{+} \to K^{*+} \mumu$ \cite{Aaij:2014pli,Aaij:2016flj,Aaij:2016cbx}, $B^0_s \to \phi \mumu$ \cite{Aaij:2015esa} and $\Lambda_b \to \Lambda \mumu$ \cite{Aaij:2015xza,Detmold:2016pkz} for di-lepton invariant masses squared in the region $[1,6]$ GeV$^2$, i.e. below the charmonium threshold. While most of the above measurements are relatively old, the very recent analyses in refs. \cite{Aaij:2016flj,Aaij:2016cbx} are entirely independent from their predecessors, yet the results show again the aforementioned pattern.

Since branching-ratio measurements are affected by large theoretical uncertainties, such as hadronic form factors, it is hard from the above data alone to draw conclusions. However, this objection can be averted by constructing ratios of branching ratios to different final-state leptons. LHCb performed the following measurements \cite{Aaij:2014ora,Aaij:2017vbb}
\beqa
\label{eq:RK}
&R_K([1, 6] {\textrm{GeV}}^2) ~\equiv~ \frac{\mc B (B^+ \to K^+ \mu^+\mu^-;~q^2 \in [1, 6] {\textrm{GeV}}^2)}{\mc B (B^+ \to K^+ e^+e^-;~q^2 \in [1, 6] {\textrm{GeV}}^2)} ~=~
0.745^{+0.090}_{-0.074}\,{\textrm{(stat)}}\pm 0.036\,{\textrm{(syst)}}~,&\nn\\
&R_{K^{*0}}([0.045, 1.1] \, {\textrm{GeV}}^2) ~=~ 0.660 ^{+0.110}_{-0.070} \pm 0.024~,&\\
&R_{K^{*0}}([1.1, 6] \, {\textrm{GeV}}^2) ~=~ 0.685 ^{+0.113}_{-0.069} \pm 0.047~,&\nn
\eeqa
where the $R_{K^{*0}}$ definition is analogous to the $R_K$ one, and $q^2$ is the invariant mass squared of the dilepton pair. Within the SM, either of the above ratios is predicted close to unity with a few-percent accuracy \cite{Bordone:2016gaq} (see also \cite{Bobeth:2007dw,Bouchard:2013mia,Hiller:2003js}). The $R_{K}$ and $R_{K^{*0}}$ measurements each imply a discrepancy between 2 and $2.6\sigma$ \cite{Aaij:2014ora,Aaij:2017vbb}, at face value signalling lepton-universality violation (LUV) beyond the SM. 

The electron-channel measurement would be an obvious culprit for either of the  $R_K$ and $R_{K^{*0}}$ discrepancies, because of bremsstrahlung and lower statistics with respect to the muon channel. On the other hand, disagreement is rather in the muon channel, see \cite{Aaij:2014pli,Aaij:2012vr} and, very recently, Ref. \cite{Aaij:2016flj}. A systematic effect in the muon channel, although not impossible, is less likely than in the electron channel, considering that muons are among the most reliable objects within LHCb.

The other $b \to s \mumu$ modes mentioned above turn out to fit a coherent picture with $R_K$ and $R_{K^{*0}}$:
\begin{itemize}

\item The $B_{s} \rightarrow \phi \mu^{+} \mu^{-}$ differential branching ratio is measured to be consistently lower than the SM prediction, in the same range $m_{\mu \mu}^{2} \in [1, 6]$ GeV$^{2}$. This was initially found in 1/fb of LHCb data \cite{Aaij:2013aln} and was afterwards confirmed by a full Run-1 analysis \cite{Aaij:2015esa}. This discrepancy is estimated to exceed 3$\sigma$ \cite{Aaij:2015esa}.

\item The $B \rightarrow K^{*} \mu \mu$ angular analysis exhibits a well-known discrepancy in one combination of the angular-expansion coefficients, known as $P_{5}'$, in, again, the same kinematic region in $q^2$ as initially measured in \cite{Aaij:2013qta}, confirmed by a full Run-1 analysis \cite{Aaij:2015oid}. The picture is further corroborated by a Belle analysis \cite{Abdesselam:2016llu,Wehle:2016yoi}, whereas the recent ATLAS and CMS measurements come at present with large error bars \cite{ATLAS_P5p,CMS_P5p}. While theoretically more debated, cf. \cite{Descotes-Genon:2013vna,Khodjamirian:2010vf,Descotes-Genon:2013wba,Lyon:2014hpa,Jager:2014rwa,Ciuchini:2015qxb} for minimal literature, this observable provides additional circumstantial support to the other discussed data.

\end{itemize}

Further interesting results come from measurements of the ratios $R(D^{(*)}) \equiv \mc B (B \to D^{(*)} \tau \nu) / \mc B (B \to D^{(*)} \ell \nu)$, but are of no direct concern in the present context. We refer the reader to \cite{Amhis:2016xyh,Gershon:2017jlb}. A generic up-to-date approach towards a common explanation of these anomalies has been given in \cite{Buttazzo:2017ixm}.

As mentioned above, the key feature of this would-be first manifestation of physics beyond the SM at collider scales is LUV. Given the somewhat unexpected nature of this conclusion, it is of utmost importance to have the largest possible number of further tests, in order to sufficiently constrain the short-distance physics responsible. A first set of such tests is the measurement of further $R_K$-like ratios, such as $R_{K^*,\,X_s,\,K_0(1430),\,\phi}$, as discussed in Ref. \cite{Hiller:2014ula}.

\medskip

In this paper we put forward one further test of LUV, namely the ratio $R_\gamma \equiv \mc B(\bsmumugamma) / \mc B(\bseegamma)$, to be properly defined in sec. \ref{sec:high-q2}. The main advantage of the above radiative modes with respect to the corresponding non-radiative counterparts is the fact that the chiral suppression factor -- especially severe in the electron channel -- is eluded by the presence of the additional photon. 
In fact, the above ratio is very close to unity, with numerator and denominator being both in the ballpark of $10^{-8}$, which should be compared with $\mc B(B_s^0 \to \mu^+ \mu^-) \simeq 3 \times 10^{-9}$ and $\mc B(B_s^0 \to e^+ e^-) \simeq 9 \times 10^{-14}$\cite{Bobeth:2013uxa}. Such relatively `large' branching ratios, and in spite of the challenges inherent in the electron channel, make these radiative modes interesting new observables for Run 2 (and beyond) of the LHC. 

As regards the sensitivity of $B^0_s \to \ell^+ \ell^- \gamma$ to beyond-SM effects, the presence of the additional photon warrants a richer short-distance structure than $B^0_s \to \ell^+ \ell^-$, because the photon lifts the chiral suppression of the purely leptonic decay \cite{Dincer:2001hu}. For this very reason, the radiative decay is promising even for lepton-flavor-violating searches \cite{Guadagnoli:2016erb,Guadagnoli:2017jcl}. As a matter of fact, the additional photon extends the sensitivity to include all the operators relevant for the previously mentioned discrepancies since the effective Hamiltonians of $B^0_s \to \ell^+ \ell^- \gamma$ and $b \to s \ell \ell$ are identical.

The two decays involved should be measured in appropriate kinematic ranges, chosen to have a sufficiently large event yield, and such that theoretical uncertainties can be kept below the level of the expected new-physics effects. (A recent search in both the $\mumu$ and $e^+ e^-$ channels was presented in \cite{Aubert:2007up}.) These two $q^2$ intervals lie respectively below and above the narrow-charmonium resonances, and are accordingly denoted as low- and high-$q^2$ regions. We disregard the narrow-charmonium region, although some of the considerations we make for the low-$q^2$ interval may one day be applicable to that region as well.

The low-$q^2$ range includes the $\phi(1020)$ resonance. We argue that the theoretical uncertainty associated to the prediction for the {\em total} $\mc B(\bsmumugamma)$ (as defined above) can be drastically reduced taking into account, for low $q^2$, its close parenthood with the measured $B_s^0 \to \phi (\to K^+ K^-) \gamma$. As concerns the high-$q^2$ region, we point out that the $q^2$-differential branching ratio is dominated by only two $B_s^0 \to \gamma$ form factors, 
to wit the vector and the axial one. As a consequence theory uncertainties do cancel to a large extent in the ratio between two different lepton channels. Keeping in mind that this region is by far dominated by the Wilson coefficients $C_9$ and $C_{10}$, such ratio provides a new, stringent test of LUV.
Incidentally, a measurement of this ratio in the low-$q^2$ region would provide a cross-check of the $\RKst$ \eqref{eq:RK} result in the lowest bin. This measurement is rather surprising, as e.m.-dipole operators, which are the dominant ones in this region, are necessarily lepton-universal. It should also be kept in mind that, due to the proximity to the kinematical threshold, a robust error assessment may be more delicate in this region \cite{Bordone:2016gaq}. The discrepancy in this bin, if confirmed, would require light new physics not describable within the effective-theory approach to be detailed in the next section.

This paper is organised as follows. In sec. \ref{sec:bsmumugamma} we discuss the effective-theory basics of the \bsmumugamma decay. Our aim is to introduce necessary notation for the ensuing discussion, on the short- vs. long-distance contributions to this decay in the two different kinematical regions considered. The two following sections, \ref{sec:low-q2} and \ref{sec:high-q2}, are devoted to a more in-depth consideration of the low- and high-$q^2$ regions, and of the main theoretical uncertainties involved. Here we put forward a number of strategies to reduce these uncertainties below the level that makes this observable a valuable new probe of the very interactions hinted at by present-day discrepancies in flavor data. In sec. \ref{sec:outlook} we collect a few ideas that, we believe, deserve further investigation. Finally, a summary of our main results and conclusions are presented in sec. \ref{sec:summary}.

\section{The \bsmumugamma decay} \label{sec:bsmumugamma}

\subsubsection*{Basic formulae}
\begin{figure}[t]
\hspace{-0.5cm}
\begin{tabular}{llll}
{\bf (a)} & {\bf (b)} & {\bf (c)} & {\bf (d)} \\
  \includegraphics[width=0.24\linewidth]{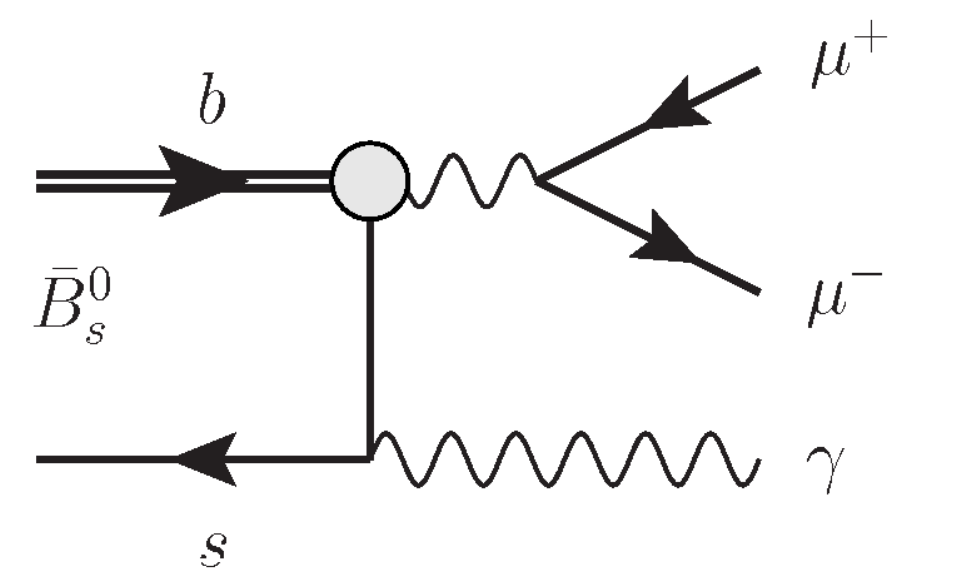} &
  \includegraphics[width=0.24\linewidth]{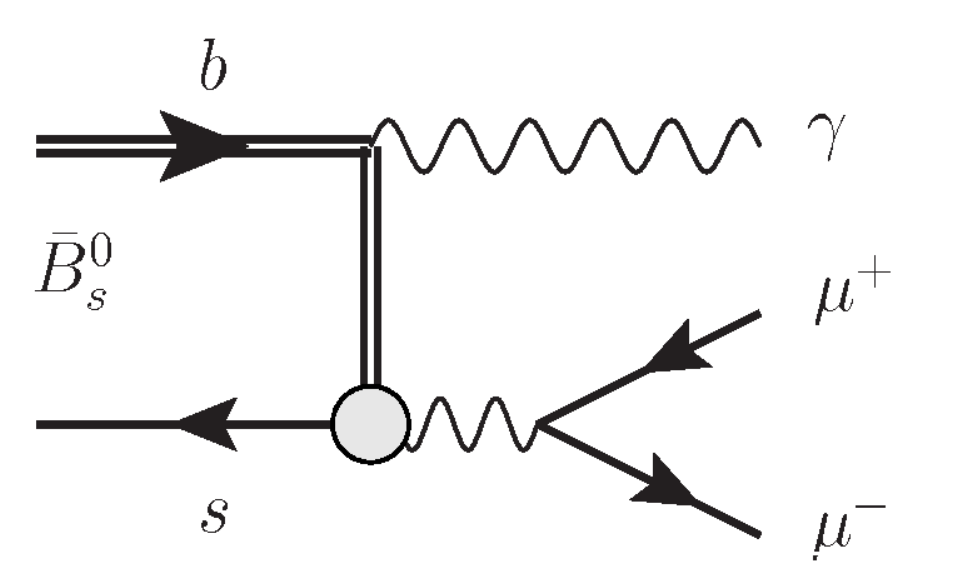} &
  \includegraphics[width=0.24\linewidth]{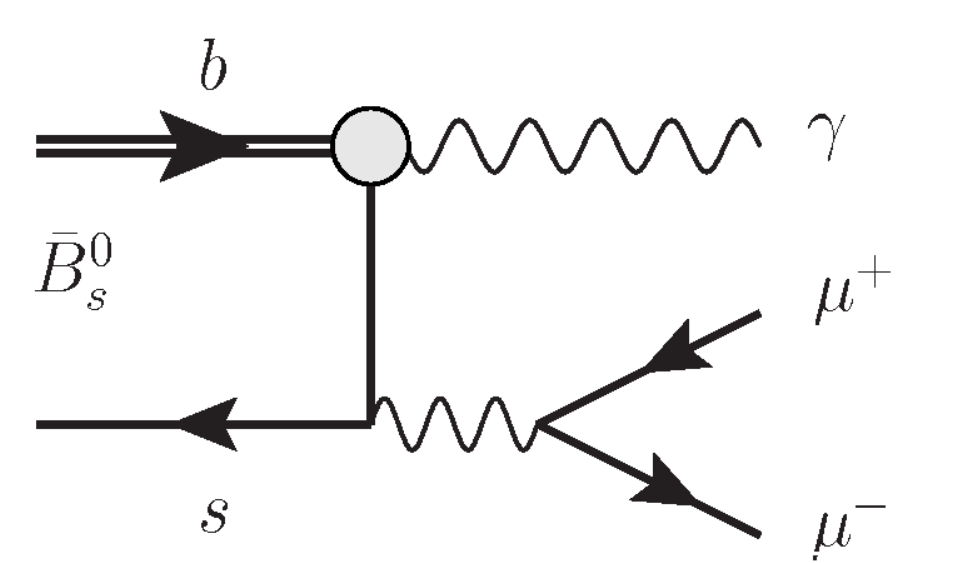} &
  \includegraphics[width=0.24\linewidth]{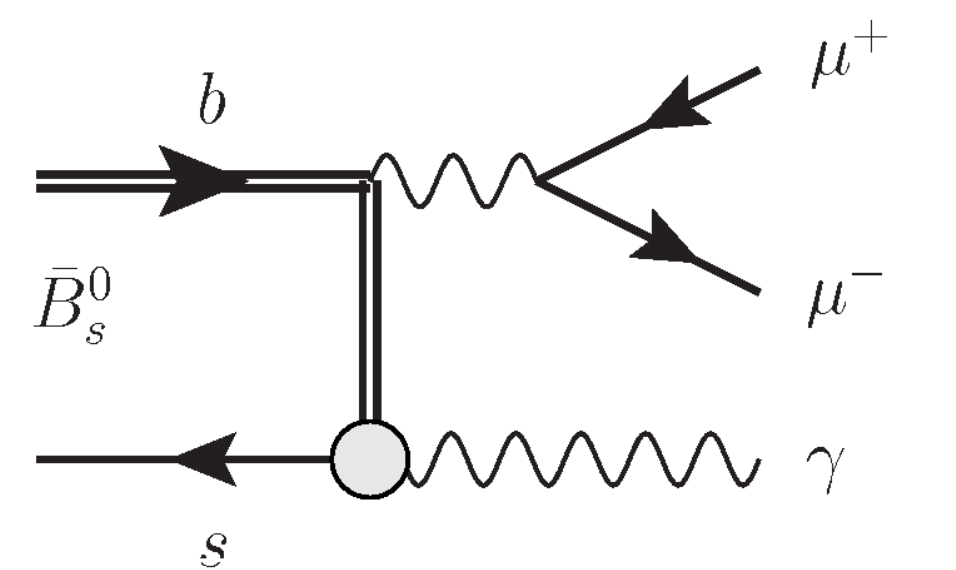} \\
[0.5cm]
{\bf (e)} & {\bf (f)} & {\bf (g)} & {\bf (h)} \\  
  \includegraphics[width=0.24\linewidth]{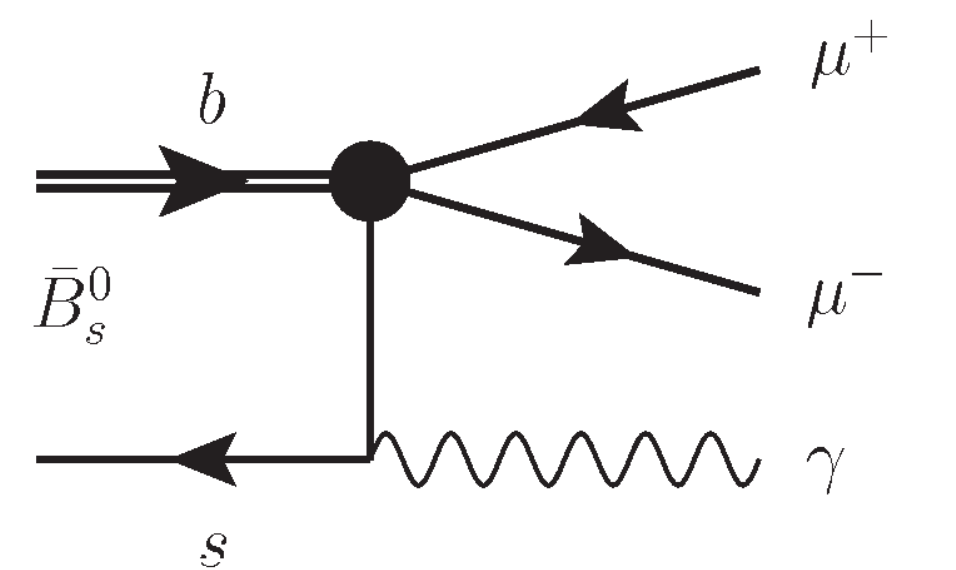} &
  \includegraphics[width=0.24\linewidth]{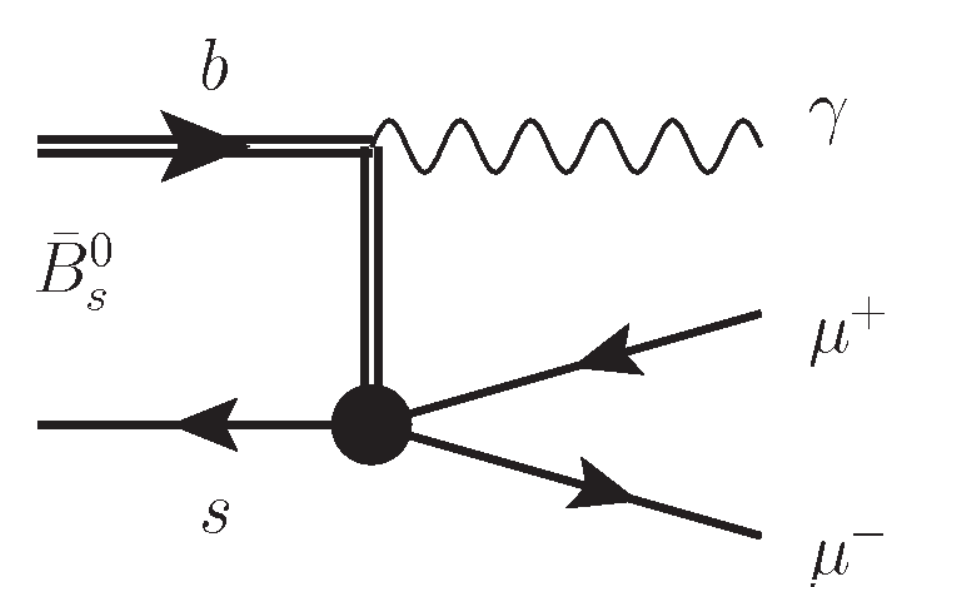} &
  \includegraphics[width=0.24\linewidth]{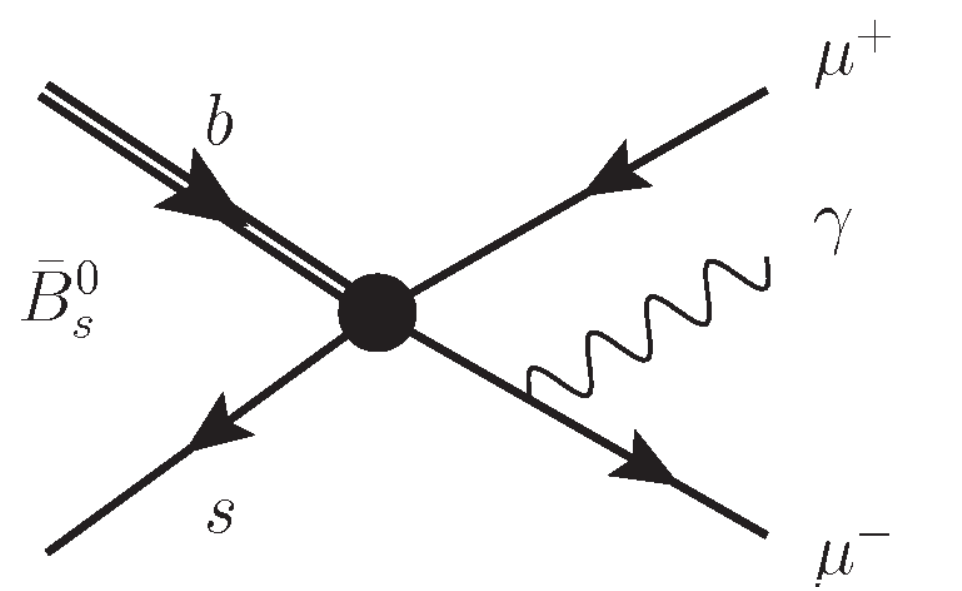} &
  \includegraphics[width=0.24\linewidth]{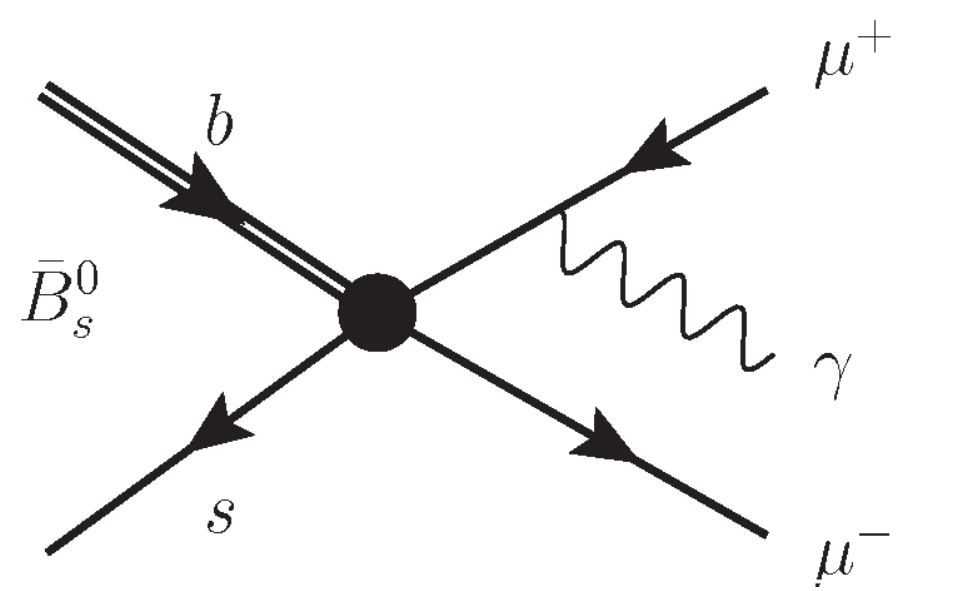}\\
\end{tabular}
  \caption{Short-distance diagrams contributing to the $\bar B^0_s \to \mumu \gamma$ process to lowest order. 
  The black and the grey circles denote the insertion of the four-fermion operators $\mc O^{(')}_{9,10}$ and respectively of $\mc O^{(')}_7$. The form factors $\FTVA(q^2,0)$ and $\FTVA(0,q^2)$ describe the diagrams $(a,b)$ and $(c,d)$ respectively. Diagrams $(e)$ and $(f)$ are described by the $\FVA(q^2)$ form factors and diagrams $(g)$ and $(h)$ encode bremsstrahlung contributions, whose hadronic matrix elements are described by the $\bar B_s^0$ decay constant.
}
  \label{fig:diags}
\end{figure}

\noindent 
The dynamics of the \bsmumugamma amplitude can be parameterised by the following $b \to s \ell \ell$ effective Hamiltonian \cite{Buchalla:1995vs,Buras:1994dj,Misiak:1992bc} 
\beq
\label{eq:Heff}
\mathcal H_{\mathrm{eff}} = \frac{4 G_F}{\sqrt 2}\left( \sum_{i=1}^2 
(\lambda_u C_i \mathcal{O}_i^u + \lambda_c C_i \mathcal  \mathcal{O}_i^c )       -\lambda_t \sum_{i=3}^{6} C_i \mathcal{O}_i   - \lambda_t \sum_{i=7}^{10} (  C_i  \mathcal {O}_i + C_i'  \mathcal{O}_i')  
 \right)~, 
\eeq
where $\lambda_i \equiv V_{is}^*V_{ib}$, with $V_{ij}$ CKM matrix elements, and $C_i$ are the Wilson coefficients. The operators that will be relevant for the rest of the discussion are defined explicitly as
\begin{alignat}{4}
\label{eq:opbasis}
& \mathcal O_1^q &\;=\;&  (\bar s_{i} \gamma_\mu q_{Lj}) (\bar q_{j} \gamma^\mu b_{Li})~,  \qquad  \qquad 
& & \mathcal O_2^q &\;=\;&  (\bar s_{i}  \gamma_\mu q_{Li}) (\bar q_{j} \gamma^\mu b_{Lj})~, \nonumber \\[0.1cm]
& \mathcal O_{7} &\;=\;& \frac{e m_b}{16\pi^2}\bar s \sigma_{\mu \nu} F^{\mu \nu} b_R~,
\qquad  \qquad 
& & \mathcal O_{8} &\;=\;& \frac{g_s m_b}{16\pi^2}\bar s \sigma_{\mu \nu} G^{\mu \nu} b_R~, 
\\[0.1cm]
& \mathcal O_{9} &\;=\;& \frac{e^2}{16\pi^2}(\bar s \gamma_\mu  b_L) (\bar \ell\gamma^\mu \ell) ~,
\qquad  \qquad
& & \mathcal O_{10} &\;=\;& \frac{e^2}{16\pi^2}(\bar s \gamma_\mu  b_L) (\bar \ell\gamma^\mu \gamma_5 \ell) ~, \nn
\end{alignat}
where $i,j$ are colour indices and the primed operators are obtained from eq. (\ref{eq:opbasis}) by the replacements \{$L \to R$, $m_b \to m_s$\}. The sign conventions for the e.m. and strong couplings of $\mathcal O_{7,8}$ are consistent with the covariant derivative $D_\mu = \partial_\mu + i e Q_f A_\mu + i g_s G_\mu$ (e.g. $Q_\mu = Q_e = -1$) and  $C^{\textrm{SM}}_{7,8} < 0$. The  $\mc O_{1-6,8}$ matrix elements give rise to long-distance (LD) contributions with distinct $q^2$ dependence and strong phases. The $\mc O^{(')}_{7,9,10}$ matrix elements are of short-distance (SD) nature, and are parameterised by form factors, see e.g. \cite{Kruger:2002gf},%
\footnote{Our notation translates into the one of Ref. \cite{Kruger:2002gf,Melikhov:2004mk} as $\FV = - F_V, \;\FA = - F_A,\;\FTV = - F_{TV}, \;\FTA = - F_{TA}$. One reason for introducing this new notation is to make contact with the $B \to V \ell \ell$ literature, where $V$ and $A$ labels refer to the polarisation of the leptons in the effective theory language. We also note that the sign of the form factors depends on the sign convention of the covariant derivative. Our covariant derivative convention, specified above, is consistent with all $\bar B_s (\bar B_u) \to \gamma$ form factors being positive (negative), since their leading contribution is proportional to the light quark charge. This can be inferred from Refs. \cite{DescotesGenon:2002mw,Ball:2003fq,Ball:2006eu} in the context of $B_u \to \ell \nu \gamma$ transitions.} of dilepton momentum transfer $q^2 = (p-k)^2$,\footnote{The photon on-shell matrix elements, for the last two expressions in \eqref{eq:ffs}, are obtained by setting $k^ 2=0$ and contracting by the polarisation vector $\epsilon^{*}_\al (k)$.}
\begin{alignat}{3}
\label{eq:ffs}
& \< \gamma(k,\epsilon) | \, \bar{s} \gamma^{\mu} (1 \pm \ga_5) b \, | \bar{B}^0_s (p) \> M_{B_s}
 &\;=\;& - e \{ \Pperp{\mu} \, \FV (q^2)  &\;\pm\;&  \Ppara{\mu} \, \FA(q^2)  \} ~, \nn \\[0.1cm]
& \< \gamma^*(k,\alpha) | \, \bar{s} i q_\nu \sigma^{\mu \nu} ( 1 \mp \ga_5) b \, | \bar{B}^0_s (p) 
\> &\;=\;& + e \{ \Pperp{\mu \al}   \, \FTV(q^2,k^2) &\; \pm\;& \Ppara{\mu \al}  \, \FTA(q^2,k^2)  
\} \;, 
\end{alignat}
and the $B^0_s$ decay constant
\beq
\label{eq:fB}
\< 0 | \, \bar{s} \, \gamma^{\mu} \gamma_{5} \, b \, | \bar{B}^0_s (p) \> 
= i p^{\mu} f_{B_s}~.
\eeq
Above
\beq
\label{eq:P}
  \Pperp{\mu \al}    =
 \varepsilon^{\mu \al \be \ga } p_\be k_\ga  \;,  \quad    \Ppara{\mu \al}   =  i  \, (   p \cdot k \, g^{\mu \al} -  p^\al  \, k^{\mu}) \;,
\eeq
where the $\varepsilon_{0123} = 1$ convention is assumed and $ \Pperp{\mu} \equiv   \epsilon^*_\al \Pperp{\mu \al}$ and analogous for the $\parallel$-direction. In practice for $\FTVA(q^2,k^2)$ either $q^2$ or $k^2$ will be zero because of the on-shell photon in the final state, cf. caption of fig.~\ref{fig:diags} and at last we note the algebraic relation 
$\FTV(0,0) = \FTA(0,0)$ \cite{Kruger:2002gf}. We comment on the theoretical status of the form 
factors in appendix \ref{app:FF}. 
We define the amplitude as
\begin{equation}
\label{eq:A}
{\cal A} \equiv \matel{\mu^+(p_1) \mu^-(p_2) \ga(k, \eps) }{(-\mathcal H_{\mathrm{eff}} )}{\bar{B}^0_s}
\end{equation}
and
derive the following SD amplitude\footnote{Cf. \cite{Melikhov:2004mk}, and \cite{Guadagnoli:2016erb} for the correct sign of the interference term in the corresponding differential width.}
\begin{alignat}{1}
\label{eq:SDamplitude}
{\cal A}_{\textrm{SD}}  = - \frac{e  \al  \la_t G_F }{\sqrt2 2 \pi} 
 \left \{ \phantom{\frac{1}{1}} \!\!\!\!   \right. &  
  \frac{2 m_b}{q^2} \left( (C_7 + \frac{m_s}{m_b} C_7') \FTVb(q^2) \Pperp{\mu} - (C_7 - 
 \frac{m_s}{m_b} C_7') \FTAb(q^2) \Ppara{\mu} \right) \bar{u}(p_2) \ga_\mu v(p_1) 
 +
  \nonumber \\
&  \frac{1}{M_{B_s}} \left( (C_{9}(q^2) + C'_{9})  \FV(q^2) \Pperp{\mu} - (C_{9}(q^2) - C'_{9}) \FA(q^2) 
\Ppara{\mu} \right) \bar{u}(p_2) \ga_\mu v(p_1) + \nonumber \\
&  \frac{1}{M_{B_s}} \left( (C_{10} + C'_{10})  \FV(q^2) \Pperp{\mu} - (C_{10} - C'_{10}) \FA(q^2) 
\Ppara{\mu} \right) \bar{u}(p_2) \ga_\mu \ga_5 v(p_1) - \nonumber \\
&  i f_{B_s} 2 m_\mu \left( C_{10} - C'_{10} \right) \bar{u}(p_2) \left( 
\frac{ \slashed{\eps}^* \slashed{p}}{t - m_\mu^2} - \frac{\slashed{p} \slashed{\eps}^*}{u - m_\mu^2} \right) \ga_5 v(p_1)~
\left. \phantom{\frac{1}{1}} \!\!\!\!  \!\! \right \},
\end{alignat}
where 
\begin{equation}
\FTVAb(q^2) =  \FTVA(q^2,0) +  \FTVA(0,q^2)  \;,
\end{equation}
takes into account diagrams $(a,b)$ and $(c,d)$ in fig. \ref{fig:diags}.

Before amending the LD part let us mention that the amplitude (\ref{eq:SDamplitude}) gives rise to a double-differential decay distribution in two of the three Mandelstam variables, or equivalently $d \Gamma( \bar B^0_s \to \ell^+ \ell^- \gamma)/d q^2 d \cos\theta$, see \cite{Kruger:2002gf,Melikhov:2004mk}. Here $\theta$ is the angle between the three-momentum vectors of the $\mu^-$ and of the photon in the dilepton center-of-mass system \cite{Kruger:2002gf}. All but the last line in eq.(\ref{eq:SDamplitude}) are $S$ and $P$-waves. It is worthwhile to mention that that the bremstrahlung contribution, being non-local in $\cos\theta$, gives rise to all higher partial waves and can in principle be filtered out by a method-of-moments analysis, as proposed in \cite{Gratrex:2015hna}.

The LD contributions involve a muon pair emitted from a photon and we may therefore parameterise the full amplitude (\ref{eq:A}) as
\begin{equation}
{\cal A} = - \frac{G_F \, \la_t}{\sqrt2} \frac{e \, \al }{2 \pi}  \left(  \frac{2 m_b}{q^2}  
\left( \APPperp(q^2)  \Pperp{\mu} - \APPpara(q^2)  \Ppara{\mu} \right) \bar{u}(p_2) \ga_\mu v(p_1) +  O(C_{9,10}^{(\prime)}) \right)  \;,
\end{equation} 
such that 
\begin{equation}
\label{eq:APPperp}
\APPb(q^2) = (C_7 \pm \frac{m_s}{m_b} C_7') \FTVAb(q^2) + 
(C_8 \pm \frac{m_s}{m_b} C_8') G_{\perp,\parallel}(q^2) + \sum_{i=1}^6 C_i L_{i \perp,\parallel}(q^2) \;.
\end{equation}
Above $G_\perp(q^2)$ and $L_i(q^2)$ stand for the chromomagnetic penguin and the four-quark operator contributions, respectively, which are the LD parts to be discussed further below.

\subsubsection*{Short- versus long-distance contributions at low and high $q^2$}

\noindent Before discussing applications in the low and high $q^2$-region in sections \ref{sec:low-q2} and \ref{sec:high-q2} respectively, we give an executive summary over the different topologies and their relevance in those regions.

The different SD contributions to the leading-order amplitude in weak interactions are displayed in fig. \ref{fig:diags}. We briefly digress on the relevance of each of these contributions in the different kinematic regions. Except for the kinematic endpoint, there are analogies with the $B \to V \ell \ell$-type decays. At very low $q^2$, the $\mc O_7$ matrix elements, diagrams $(a)$-$(d)$, dominate, because of the proximity of the photon pole. This holds true in the region of the $\phi$(1020) resonance. With regard to the $\mc O_7$ diagrams it is useful to distinguish between diagrams $(a,b)$, where the off-shell photon is the one emitted from the penguin, and its momentum dependence is described by the form factors $\FTV(q^2,0)$ and $\FTA(q^2,0)$, and diagrams $(c,d)$, in which the off-shell photon momentum probes the bottom and strange quark currents, and its momentum dependence is described by the form factors $\FTV(0, q^2)$ and $\FTA(0, q^2)$. For $q^2 \simeq M_\phi^2$ the subprocess $B_s^0 \to (\phi \to \mu\mu) \gamma$ dominates, which is of crucial importance for this paper. Concerning the $\mc O_{9,10}$ diagrams,  it is in turn helpful to distinguish between diagrams $(e,f)$ and $(g,h)$, where the photon is radiated from the strange or bottom quark, and the final state leptons, respectively. The latter are of the bremsstrahlung (also referred to as final-state radiation, FSR) type, are described by the $B^0_s \to$ vacuum matrix element (\ref{eq:fB}), and are dominant close enough to the kinematic endpoint.

\begin{figure}[t]
\begin{tabular}{ll}
{\bf (i)} & {\bf (j)} \\
 \hspace{0.5cm} \includegraphics[width=0.40\linewidth]{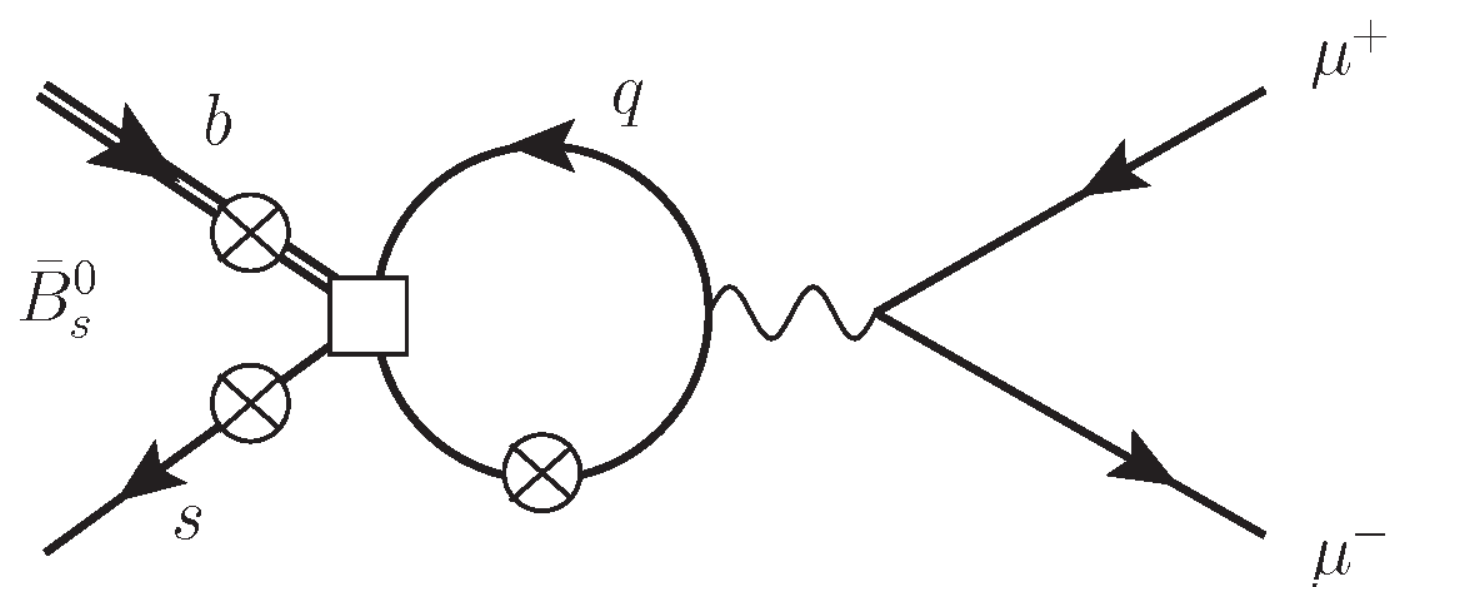} \hspace{1cm} & 
  \includegraphics[width=0.40\linewidth]{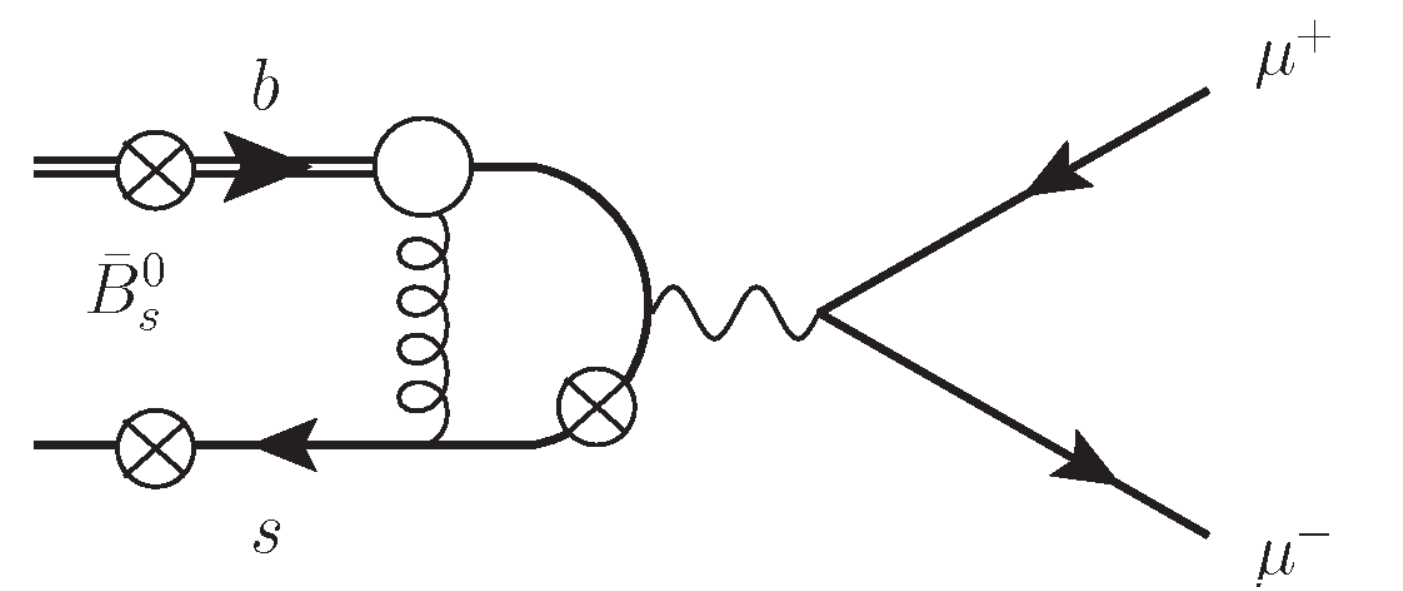} \hspace{0.5cm} \\
\end{tabular}
  \caption{Long-distance diagrams contributing to the $\bar B^0_s \to \mumu \gamma$ process to lowest order. The empty square or circle denote the insertion of one of the four-quark operators in eq. (\ref{eq:Heff}) and, respectively, of the operator $\mc O_8$. The symbol $\otimes$ denotes all the possible ways of attaching an on-shell photon, with exclusion of bremsstrahlung.}
  \label{fig:LDdiags}
\end{figure}

Next we would like to  discuss the leading LD topologies. The latter, see fig. \ref{fig:LDdiags}, are the four-quark contributions, which we shall loosely refer to as weak annihilation (WA) following the terminology in \cite{Melikhov:2004mk},%
\footnote{In $B \to V \ell \ell$, one distinguishes between quark loops and WA for four quark operators. This terminology will become relevant when discussing the $B_s^0 \to \phi( \to \mu \mu) \gamma$ subprocess.} and the contribution from the chromomagnetic $\mathcal{O}_8$.
LD contributions are relevant at very low $q^2$ due to the $1/q^2$ enhancement from the virtual photon emitting the lepton pair, and in the respective resonance region(s). 

WA has been computed for low $q^2$ in \cite{Melikhov:2004mk}, at leading twist-2 (i.e. perturbative photon) and in the limit of massless up- and charm-quarks. We note that this contribution can also be obtained from the WA amplitude for the $B^0_d \to \gamma \gamma$ decay, as computed in Ref. \cite{Bosch:2002bv}, including its full up- and charm-quark mass dependence. In the notation of eq. \eqref{eq:APPperp} this contribution reads
\beq
L_{1\perp}  ~=~ - \frac{8}{3} \frac{f_{B_s}}{m_b}  \frac{1 }{\la_t} \left( \la_u g(z_u) + \la_c g(z_c) \right) \;, 
\quad L_{2\perp} = \frac{1}{3}L_{1\perp} \;,
\eeq
with $ z_i \equiv m_i^2 / m_b^2$ and the $g(z)$ function defined in \cite{Bosch:2002bv}.\footnote{Note that the definitions of $\mc O_{1,2}$ in \cite{Bosch:2002bv} are interchanged with respect to our notation in eq. (\ref{eq:Heff}), which follows \cite{Buchalla:1995vs,Buras:1994dj,Misiak:1992bc}. We use $C_1(m_b) = -0.278$, $C_2(m_b) = 1.123$.} We include this contribution in our analysis. In this work we disregard the corresponding contributions from 4-quark operators other than $\mc O_{1,2}$ (``quark loops'') because of their small Wilson coefficients.

The chromomagnetic matrix element of $\mathcal{O}_8$ is unknown but we can expect it to be small, as is the case for $B \to V \ell \ell$ \cite{Dimou:2012un}, in part because of, again, the small Wilson coefficient.

As concerns the $O(\alpha_s)$ corrections only partial information is available. We first discuss the low-$q^2$ region, which can be described by light-cone physics. The discussion of the different contributions necessitates the choice of a formalism. We discuss them in the light-cone sum rule (LCSR) approach with photon distribution amplitude (DA). Then the gluon can connect to the vertex and the spectator quark or to the photon-DA itself. None of these contributions are known, but we can build on our knowledge from $B \to V \ell \ell$ decays. The vertex correction has been shown to factorise into form factors times a loop function, known, from the inclusive $b \to s \ell \ell$ \cite{Asatryan:2001zw}, in leading order (LO) in $1/m_b$ within QCD factorisation \cite{Beneke:2001at,Bosch:2001gv}. This contribution is sizeable because it is, unlike the $O(\alpha_s^0)$ part, not large-$N_c$ suppressed. Hard spectator corrections can be expected to be small, judging from their $1/m_b$ contributions in $B \to V \ell \ell$ \cite{Beneke:2001at}. The contribution of the gluon emitted into the photon DA is also relatively small \cite{Ball:2006eu,Muheim:2008vu} and we expect the same to hold for the case at hand. In summary, the known vertex corrections in the $1/m_b$ limit might well be the most important contributions for the low-$q^2$ region (i.e. below the narrow charmonium resonances), which we therefore include.\footnote{Further corrections, e.g. corrections beyond the $1/m_b$-limit which do not inherit the polarisation structure of the SD form factors,  can be added once the $\bsmumugamma$ decay enters a large-statistics regime. Such corrections are for example relevant in the high-statistics $B \to K^* \mu \mu$ channel \cite{Blake:2017wjz} where they might explain the $P_5'$ angular anomaly  \cite{Blake:2017wjz} and are of importance for the search of right-handed currents e.g. \cite{Muheim:2008vu}.}

At low-$q^2$ the $\bsmumugamma$ decay is dominated by the subprocess $B^0_s \to \phi \gamma$, measured to $10\%$ accuracy and deemed to further improve. This point is discussed and exploited in section \ref{sec:low-q2} below.

In the high-$q^2$ region, above the two narrow charmonium resonances $J/\psi$ and $\psi(2S)$, the form factors associated with the genuinely SD operators, in particular $\mc O_{9,10}$, are formally dominant over the matrix elements of LD operators\cite{Grinstein:2004vb,Beylich:2011aq}, as the latter are suppressed by $3$ powers of $m_b \simeq \sqrt{q^2}$. This formal suppression may however be overruled by the broad charmonium resonances. In $B \to K \ell \ell$, estimates using naive factorisation \cite{Beylich:2011aq} are not sufficient to account for 
the effect \cite{Aaij:2013pta,Lyon:2014hpa}. 
 In section \ref{sec:high-q2} we will build on this knowledge and introduce a suitable resonance function to judge its impact on the there defined $R_\ga$ observable, sensitive to LUV new physics.

\section{The low-$q^2$ region} \label{sec:low-q2}

\subsubsection*{Modeling the $\phi$ and higher resonances}

In the low-$q^2$ region the branching fraction is dominated by the $\phi$(1020) resonance. For example, the $q^2$ interval between $2 m_\mu^2$ and 1.7 GeV$^2$ (corresponding to $\sh = 0.06$) contributes  75$-$80\% of the total branching ratio, defined by cutting out the narrow charmonium interval $[8.6, 15.8]\,{\textrm{GeV}}^2$ (corresponding to $\sh \in  [0.30, 0.55]$).\footnote{\label{foot:PHOTOS}Close to the endpoint region $\sh \approx 1$ one must take into account the effect of soft resummed bremsstrahung emission \cite{Isidori:2007zt,Buras:2012ru}. This step is performed by an experimental Monte Carlo, using e.g. PHOTOS \cite{Davidson:2010ew}. This approach is more trustworthy than a theoretical calculation, because, for example, it is able to account for photon efficiencies.}
This singles out the subprocess $B_s^0 \to (\phi  \to \mu \mu) \ga$ which in the SM is dominated by the e.m.-dipole interactions and therefore offers a sensitive probe of the $C_7$ and $C'_7$ Wilson coefficients as well as of other interactions, mediated by light new particles, and as such beyond the effective-theory picture. 

The reduced amplitude \eqref{eq:APPperp} can be written as an $n$-times subtracted dispersion relation
 \begin{eqnarray}
\label{eq:disprel}
\APP_\iota (q^2)  &\;=\;&  \sum_{k=0}^{n-1} \frac{1}{k!}  \APP_\iota^{(k)} (s_0) (q^2-s_0)^k  \nonumber \\[0.1cm]
 &\;+\;& 
(q^2-s_0)^{n}  \frac{1}{2 \pi i}   \int_{\textrm{cut}}^\infty \frac{ \textrm{Disc}[ \APP_\iota (s)] ds}{(s-q^2-i0)(s-s_0)^n} \;,
\end{eqnarray}
where hereafter $\iota = \perp,\parallel$, $ \textrm{Disc} [f](s)\equiv f(s+i0)-f(s-i0) $
 and it is assumed that the the only singularity in the physical sheet of the $q^2$ plane runs over the real axis. 
We emphasise that whereas a dispersion relation is possible for the entire $\APP_\iota (q^2)$ amplitude it is only indispensable for the part of the amplitude which is not well described by perturbative theory because of resonant behaviour. In the narrow-width approximation
\beq
\label{eq:ImF=phi}
\frac{1}{2 \pi i}\textrm{Disc}[ \APP_\iota (s)] = -  \delta(s- M_\phi^2) M_\phi f_\phi  {\cal \hat{A}}^{\bar{B}_s^0 \to \phi \gamma}_{\iota} + \dots
\eeq
where  
$ \langle \phi | \bar{s}\gamma_\mu s|0\rangle  =  f_\phi M_\phi \epsilon^*_\mu$ 
is the $\phi$ decay constant
and the dots stand for higher resonance states such as the $\phi(1680)$ and other $KK$-continuum, to be commented later on. In eq. (\ref{eq:ImF=phi}) 
${\cal \hat{A}}^{\bar{B}_s^0 \to \phi \gamma}_{\iota}$ is the formal analogue of 
$\APP^{(\bar{B}_s^0 \to \mu \mu \ga)}_\iota$ \eqref{eq:APPperp},
\begin{equation}
\label{eq:A_V,A}
{\cal \hat{A}}^{\bar{B}_s^0 \to \phi \gamma}_{\perp,\parallel}  =  
\left( C_7  \pm \frac{m_s}{m_b} C_7'\right) T^{\bar{B}_s^0 \to \phi}_{\perp,\parallel}(0) + 
\left( C_8  \pm \frac{m_s}{m_b} C_8'\right) G^{\bar{B}_s^0 \to \phi}_{\perp,\parallel}(0)
+ \sum_{i=1}^6 C_i L^{\bar{B}_s^0 \to \phi}_{i \perp,\parallel} \;.
\end{equation}
In  more standard notation, e.g. \cite{Straub:2015ica}, the form factors are denoted by 
$T^{\bar{B}_s^0 \to \phi}_{\perp,\parallel}(0) = 2 T^{\bar{B}_s^0 \to \phi}_1(0) = 2 T^{\bar{B}_s^0 \to \phi}_2(0) $
and the equality of the polarisations is the analogue of the previously 
mentioned algebraic relation below eq. \eqref{eq:P}.

To exemplify \eqref{eq:disprel} with $s_0 = 0$ and zero and one subtraction ($n = 0,1$ respectively) one obtains beyond the narrow width approximation 
\begin{alignat}{4}
\label{eq:FTVTA0q2}
\APP_\iota (q^2) =   
\left\{
\begin{array}{ll}   \frac{f_\phi M_\phi  {\cal \hat{A}}^{\bar{B}_s^0 \to \phi \gamma}_{\iota} }{q^2 - M_\phi^2 + i M_\phi \Gamma_\phi} + \dots  \phantom{go left please please}& ~~[n = 0]    \\[0.2cm]
\APP_\iota (0) +   \frac{q^2}{M_\phi^2}   \, \frac{f_\phi M_\phi  {\cal \hat{A}}^{\bar{B}_s^0 \to \phi \gamma}_{\iota} }{q^2 - M_\phi^2 + i M_\phi \Gamma_\phi}~ + ~\dots, & ~~ [n=1]
\end{array} \right. \;,  
\end{alignat}
with $\Gamma_\phi$ being the decay width 
of the $\phi$ meson. The $n=1$ version of this expansion  corresponds to the one given in \cite{Melikhov:2004mk} in the approximation $ {\cal \hat{A}}^{\bar{B}_s^0 \to \phi \gamma}_{\perp,\parallel}   = 2 T^{\bar{B}_s^0 \to \phi}_1(0) \times C_7$ with the identification $T_1^{\bar{B}_s^0 \to \phi}(0) = -g_+^{\bar{B}_s^0 \to \phi}(0)$ \cite{Melikhov:2004mk}.

Let us briefly discuss the status of knowledge of the various contributions entering 
$\bar{B}_s^0 \to \phi \gamma$. The form factor is known most precisely from LCSR, yielding \cite{Straub:2015ica}
\beq
\label{eq:g+LCSR}
T_1^{\bar{B}^0_s \to \phi}(0) =  0.309 \pm 0.027
\eeq
at twist-4 tree level and twist-3 $ O(\al_s)$  which updates the analysis \cite{Ball:2004rg,Ball:2006eu} in input parameters and a twist-4 tree-level contribution.\footnote{Another approach is based on relativistic quark models, with meson wave-functions constrained by leptonic decay constants. Predictions for $T_1^{\bar{B}^0_s \to \phi}(0)$ range between $0.38$ \cite{Melikhov:2000yu} and $0.28$ \cite{Dmitri_private} depending on the parameters used, indicating a degree of model dependence.} The $\mathcal O_8$ and four quark topologies are known in the $1/m_b$-limit \cite{Beneke:2001at,Bosch:2001gv} and in LCSR which does rely on an $1/m_b$-expansion from Refs. \cite{Dimou:2012un}  and \cite{Lyon:2013gba} respectively.

\subsubsection*{Using $B^0_s \to \phi \gamma$ data}

An alternative and possibly more effective strategy towards improving the prediction 
\eqref{eq:A_V,A} is to extract the amplitudes ${\cal \hat{A}}^{\bar{B}_s^0 \to \phi \gamma}_{\perp,\parallel}$ from experiment and then use them as probes of the interference components in the \bsmumugamma rate. This approach is promising since the branching ratio \cite{Olive:2016xmw}
\beq
\label{eq:Bsphigamma_exp}
\mc B(\bar{B}^0_s \to \phi \gamma) = (3.52 \pm 0.34) \times 10^{-5}
\eeq
is known to $10\%$ accuracy. An update including the entire Run-1 dataset of 3/fb is in progress, and further updates from Run-2 data will follow up. Hence the statistical component of the error on this measurement -- about half of the error quoted in eq. (\ref{eq:Bsphigamma_exp}) -- will decrease steadily.\footnote{On the other hand, as this measurement, like any $B_s$ mode, is obtained from a ratio with respect to a suitable $B_d$ mode, its error is eventually limited by the uncertainty on the ratio of the $B_d$ and $B_s$ hadronisation fractions in $pp$ collisions, $f_s/f_d$, currently of about 7\% \cite{Aaij:2011jp}.} 
As a consequence one can expect to extract the amplitudes at the $5\%$ level, which compares favourably to a theory error which is above $10\%$.

There is a complication though, in that $\mc B(\bar{B}^0_s \to \phi \gamma) \sim \left( |{\cal \hat{A}}^{\bar{B}_s^0 \to \phi \gamma}_{\perp}|^2 + |{\cal \hat{A}}^{\bar{B}_s^0 \to \phi \gamma}_{\parallel}|^2 \right)$ and does not provide enough information for two complex amplitudes. This situation can be improved through theoretical knowledge and related observables in this channel. First in the SM the amplitude is dominated by the form-factor component \cite{Beneke:2001at,Bosch:2001gv,Ball:2006eu,Muheim:2008vu,Dimou:2012un,Lyon:2013gba}. From these references it seems that the imaginary part does not exceed $10\%$, which bounds the strong phase to just below $6^\circ$. We also note that for $b \to s$ transitions $\la_u$ is negligible, which renders the discussion of the weak phase unimportant. Further knowledge on the amplitudes in terms of information about polarisation and phase may be obtained from direct and time-dependent CP asymmetries, see e.g. the formulae in \cite{Muheim:2008vu}. Actually, for the latter LHCb has reported a first value ${\cal A}_\Delta \simeq -0.98 (50)(20)$ \cite{Aaij:2016ofv} with a large uncertainty but also with a large deviation from the SM prediction  ${\cal A}_\Delta \simeq 0.047 (28)$ \cite{Muheim:2008vu}.
 
\begin{figure}[t]
\begin{center}
  \includegraphics[width=0.48\linewidth]{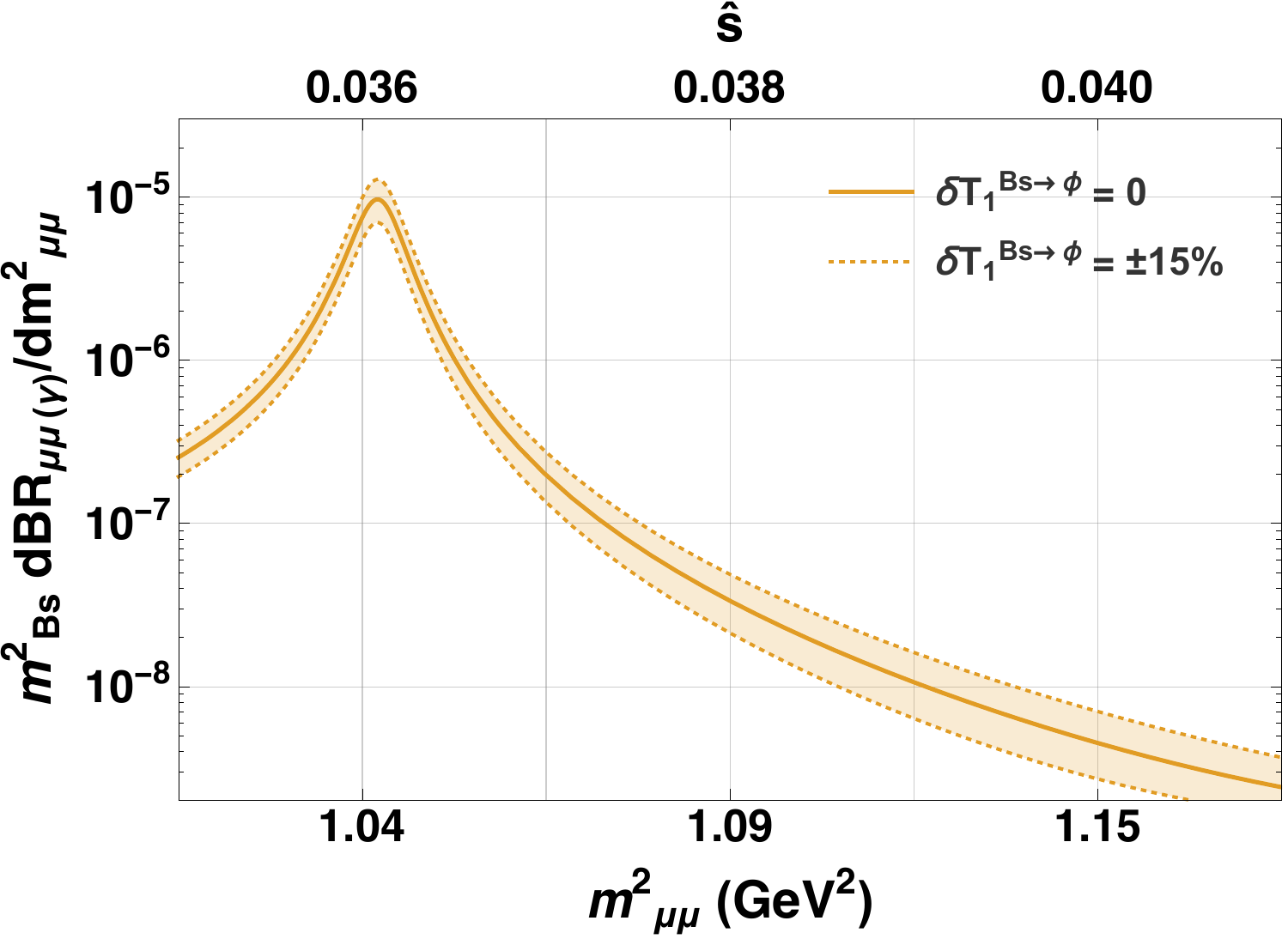}
  \includegraphics[width=0.48\linewidth]{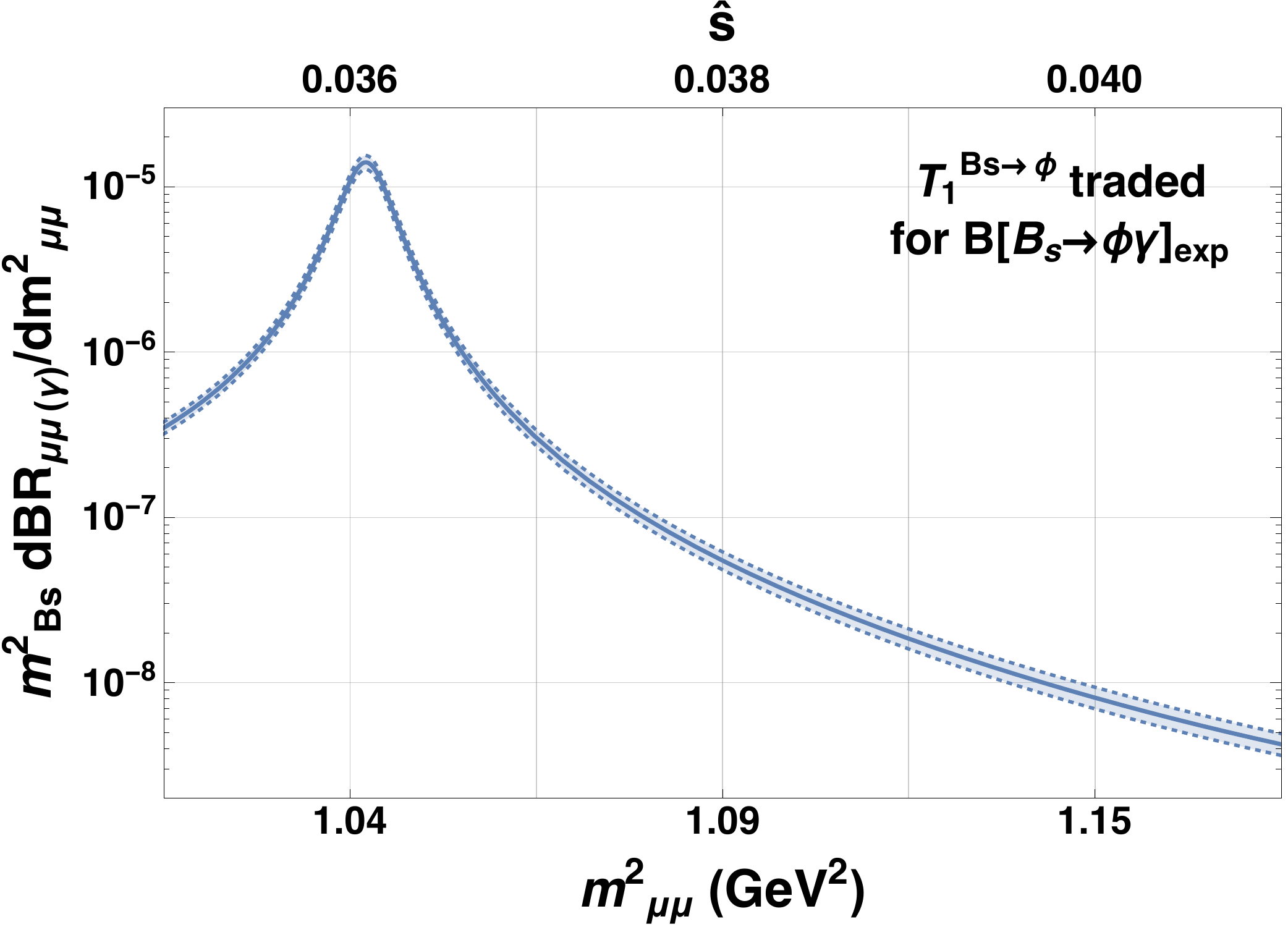}
  \caption{Differential branching ratio for \bsmumugamma around the $\phi(1020)$ resonance, with (left panel) $g_+^{B_s^0 \to \phi}$ from eq. (\ref{eq:g+LCSR}), and with (right panel) the $\phi$-peak region constrained from the {\em present} experimental measurement of $\mc B(B^0_s \to \phi \gamma)$.}
  \label{fig:phi_region}
\end{center}
\end{figure}

In fig. \ref{fig:phi_region} we display a simplified study assuming that both the $\bar B^0_s \to \phi \gamma$ and the non-resonant part of the spectrum are form-factor dominated. This figure shows the impact on the prediction of the low-$q^2$ spectrum of trading the main form-factor uncertainty, that on $\FTVA(0,q^2)$ for the measured $\bar B^0_s \to \phi \gamma$ branching ratio. Specifically, we show the $d \mc B(\bar B_s^0 \to \mumu \gamma)/d q^2$ spectrum with $T_1^{\bar{B}^0_s \to \phi}(0)$ from eq. (\ref{eq:g+LCSR}) with an error of 15\% (left panel), and (right panel) with $T_1^{\bar{B}^0_s \to \phi}(0)$ traded for eq. (\ref{eq:Bsphigamma_exp}). The 15\% error on the left panel is indicative, but can be motivated using a theoretical or an experimental argument. On the one hand, since the form-factor error is around $10\%$ and the LD part is likewise around $10\%$, one can understand the $15\%$ as a Gaussian average. Alternatively, the central value in eq. (\ref{eq:g+LCSR}) implies $\mc B(\bar B^0_s \to \phi \gamma)_{\textrm{form factor}} = 2.7 \times 10^{-5}$, which is about 30\% lower than the central value in eq. (\ref{eq:Bsphigamma_exp}), again justifying an error of about $15\%$.
As concerns the systematic error inherent in the choice of the form-factor parameterisation \cite{Kruger:2002gf}, it can only be estimated and included once an alternative evaluation, possibly from first principles, is available. More considerations on this important aspect are presented in appendix \ref{app:FF}.
The reduction in the error in the right with respect to the left figure displays the potential gain of the method, which has further potential for improvement with more statistics. On the systematic side, one needs to go beyond form-factor dominance.

\subsubsection*{Additional resonances}

In order to make the discussion following eq. (\ref{eq:ImF=phi}) more transparent, we have restricted ourselves to the case of one single resonance, the $\phi(1020)$. At this stage it is very important to assess the potential impact on the prediction of the low-$q^2$ $\bsmumugamma$ spectrum of other resonances -- most notably the $\phi(1680)$, to be denoted as $\phi^\prime$ hereafter. The $B_s^0 \to \phi'$ form factor can be estimated by scaling by the decay constants, 
\beq
T_1^{B_s^0 \to \phi'} \simeq f_{\phi^\prime} / f_{\phi} \, T_1^{B_s^0 \to \phi}~.\nn
\eeq
This is the case since the decay constant is the first term in the partial-wave expansion of the vector meson distribution amplitude. This implies $\mc B(B \to V \gamma) \propto |T^{B \to V}_1|^2 \propto f_V^2 $.  As we are unaware of an evaluation of $f_{\phi^\prime}$ we resort to $K^*$-meson data, assuming $f_{\phi^\prime} / f_{\phi} \simeq f_{K^{*\prime}} / f_{K^*}$. From the ratio between $\mc B(B \to K^*(1410) \gamma)$ and $\mc B(B \to K^*(892) \gamma)$ data \cite{Olive:2016xmw}, suitably corrected for the relevant kinematic factors, we get $f_{\phi^\prime} / f_{\phi} \simeq 0.86$. We note that this value is encouragingly close to $f_{\rho^\prime} / f_{\rho} = 0.875$ from Ref. \cite{Bakulev:1998pf} using non-local condensate sum rules. 

For such a potentially large coupling, it is clear that including or not the $\phi^\prime$ would considerably alter the prediction of the $\bsmumugamma$ spectrum at low $q^2$, hence of the $\bsmumugamma$ branching ratio as a whole. However, the large width of the $\phi'$ ($\Gamma_{\phi^\prime} \simeq 35 \times \Gamma_{\phi}$ \cite{Olive:2016xmw}) turns out to suppress the $\phi'$ contribution to the $\bsmumugamma$ spectrum to be a below-1\% correction to the total branching ratio. Needless to say, our argument can put on more solid grounds with data on $\mc B(B_s^0 \to \phi^\prime \gamma)$, to be measured in a statistically favorable $\phi^\prime$ decay mode, for example $\phi' \to K \bar K^*(892)$. Such data are not yet available at present \cite{Olive:2016xmw} and we would like to emphasise their interest, not only to robustly assess the systematics due to the $\phi'$, but also, potentially, for interference studies. We remark in fact that a large phase in the $\phi'$ Breit-Wigner would entirely cancel the suppression due to $\Gamma_{\phi'} \ll
\Gamma_{\phi}$.

A final remark is in order on the charmonium region. Attempting a description of this region with an approach similar to eq.~(\ref{eq:FTVTA0q2}) is, in principle, possible, because the $J/\psi$ and $\psi(2S)$ resonances are sufficiently narrow. On the other hand, the required radiative branching ratios are, again, not yet measured. We also remark that, at variance with the low-$q^2$ range, in this region the SD dynamics is dominated by the $\mc O_{9,10}^{(\prime)}$ operators, that one can more cleanly extract from the region $\sh > 0.55$ to be discussed next.

\section{The high-$q^2$ region} \label{sec:high-q2}

\subsubsection*{The ratio $r_\gamma$}

We next consider the part of the spectrum above the narrow-charmonium resonances, $\sh$ $\gtrsim 0.55$. As concerns the theoretical error in this region, the first consideration to be made is that the by far largest contributions come from just two sets of terms, those proportional to $\FV^2$ or to $\FA^2$, because of $C_{9,10}$-dominance,  followed by the impact on broad charmonium resonances which we address in a later section. Terms proportional to all other 
form-factor combinations have an impact that numerically does not exceed a few percent. Furthermore the existing theoretical predictions of $\FV$ and $\FA$, as well as their associated errors, are partly correlated.

\begin{figure}[t]
\begin{center}
  \includegraphics[width=0.64\linewidth]{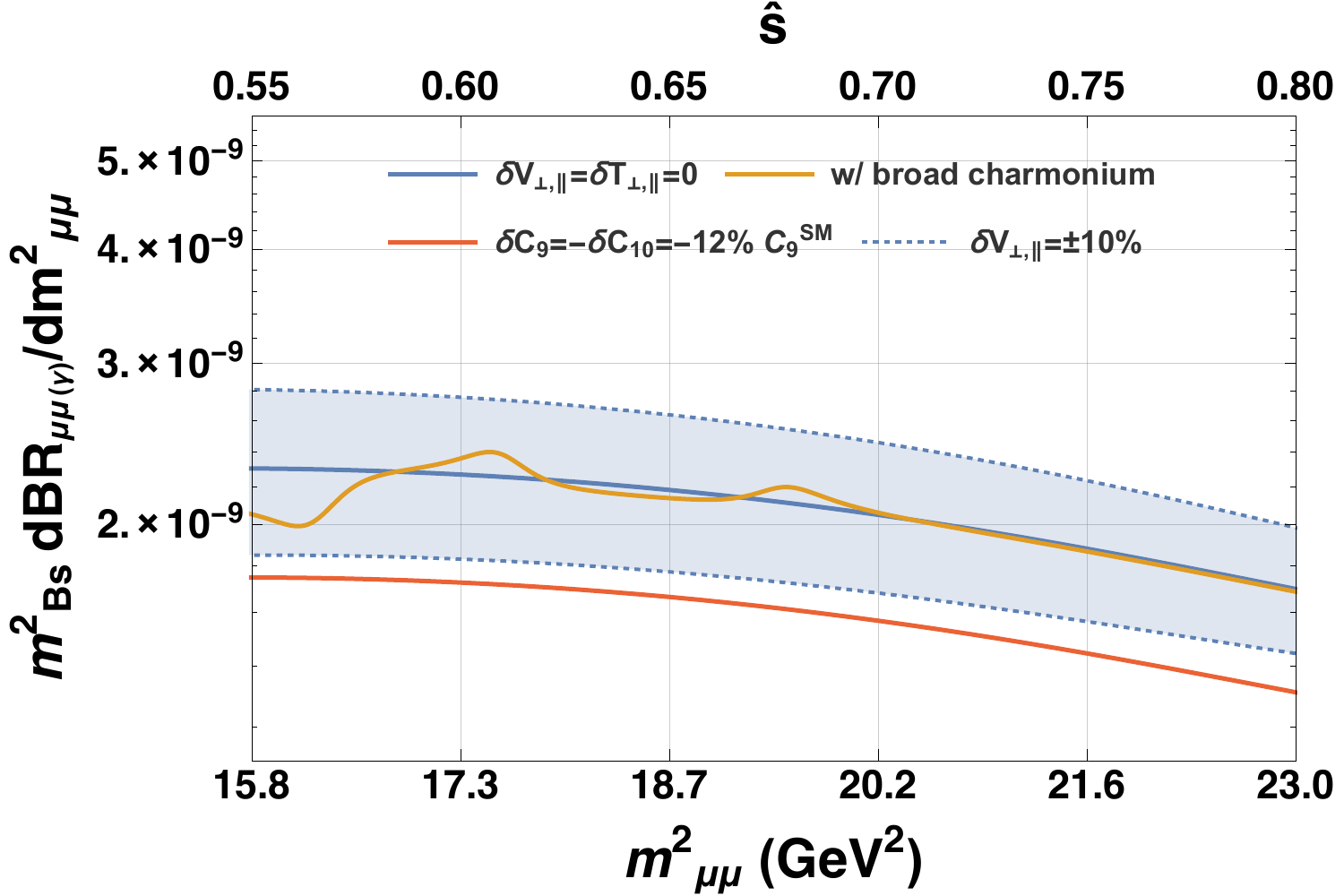} \vspace{0.5cm} \phantom{ciao}
  \includegraphics[width=0.64\linewidth]{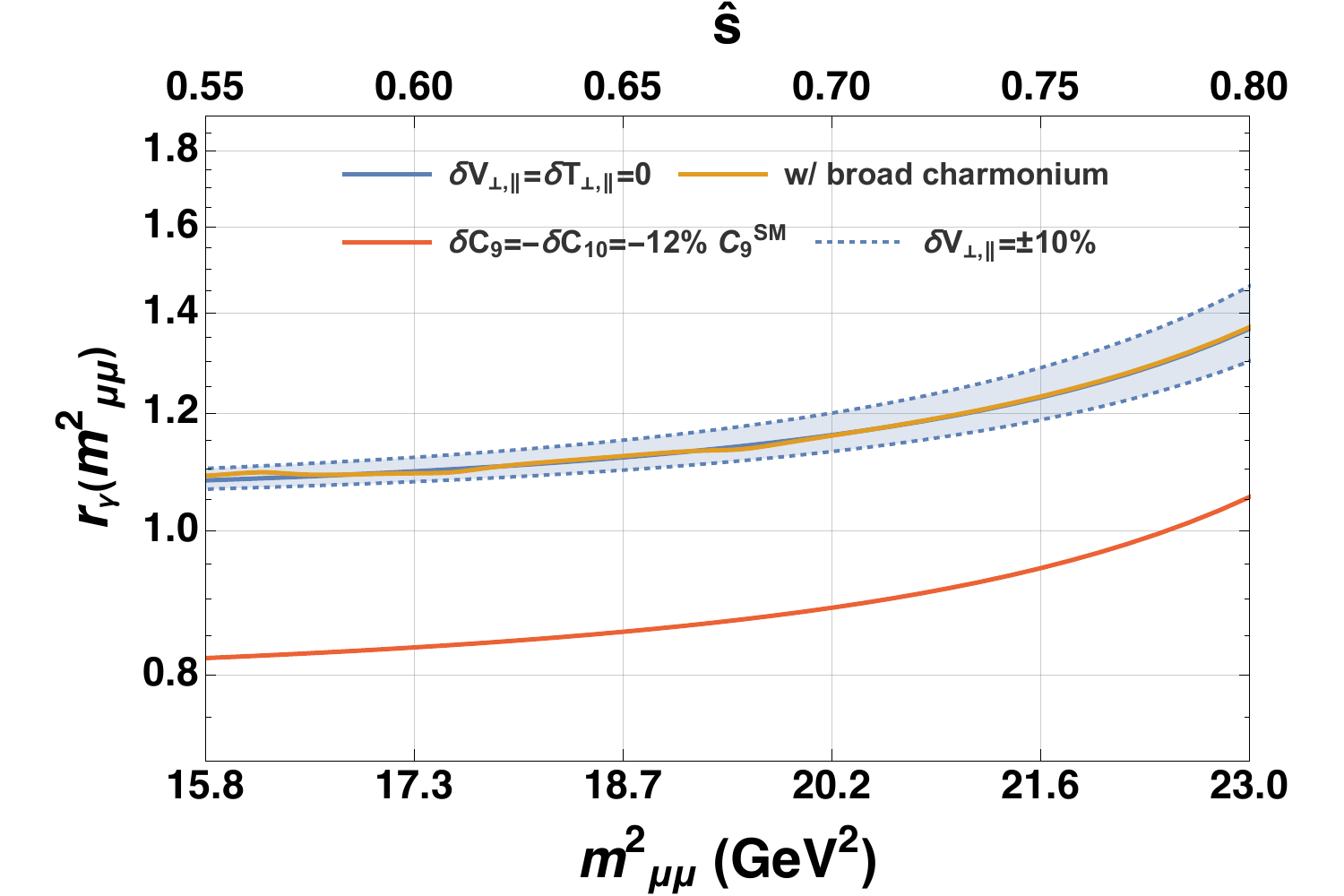}
  \caption{Comparison between (left panel) the theoretical error in the \bsmumugamma spectrum and (right panel) the corresponding error on $r_\gamma$ as defined in eq. (\ref{eq:r_gamma}).}
  \label{fig:r_gamma}
\end{center}
\end{figure}
From these considerations it is clear that the ratio of the $B^0_s \to \ell^+ \ell^- \gamma$ differential branching ratios between two different lepton channels offers a potentially much cleaner quantity than the two branching ratios considered separately. As such, this ratio provides a valuable test of lepton universality violation, in a channel devoid of final-state hadrons. More specifically, let us consider the following quantity

\beq
\label{eq:r_gamma}
r_\gamma(q^2) ~\equiv~ \frac{d \mc B (B^0_s \to \mumu \gamma) / dq^2}{d \mc B (B^0_s \to e^+ e^- \gamma) / dq^2}~,
\eeq
as well as
\beq
\label{eq:R_gamma}
R_\gamma(q_1^2, q_2^2) ~\equiv~ \frac{\int_{q_1^2}^{q_2^2} d q^2 ~ d \mc B (B^0_s \to \mumu \gamma) / dq^2}{\int_{q_1^2}^{q_2^2} d q^2 ~ d \mc B (B^0_s \to e^+ e^- \gamma) / d q^2}~,
\eeq
where we choose $q_1^2 / M_{B_s}^2 = 0.55$ (corresponding to $q_1^2 = 15.8$ GeV$^2$), i.e. somewhat above the $\psi(2S)$ resonance, and $q_2^2 / M_{B_s}^2 = 0.8$ ($q_2^2 = 23.0$ GeV$^2$) due to bremsstrahlung in the $\mumu$ channel, as explained below. The ratio $r_\gamma(q^2)$ has the following properties:

\begin{itemize}

\item Among the Wilson coefficients appearing in the Hamiltonian (\ref{eq:Heff}), the by far largest SM contributions are those from $C_{9,10}$, and the largest sensitivity is correspondingly to $C_{9,10}^{(\prime)}$. The ratio $r_\gamma$ therefore offers a further test of the very same new-physics contributions that would be responsible for $R_K$ and $\RKst$.

\item The radiative branching ratios for the $\mumu$ and for the $e^+ e^-$ channels appearing in $r_\gamma$ are very close to each other, and not hierarchically different, as in the corresponding non-radiative decays. In fact, either of $\mc B(\bsmumugamma)$ and $\mc B(\bseegamma)$, integrated over the whole $q^2$ range, are of the order of $10^{-8}$. We note explicitly that such rate, in the case of the $e^+ e^-$ channel, amounts to an enhancement over the non-radiative branching ratio of about 5 orders of magnitude.

\item As mentioned above, both numerator and denominator on the r.h.s. of eq. (\ref{eq:r_gamma}) are dominated by terms proportional to $\FV^2$ or $\FA^2$ for $\sh \in [0.55, 0.8]$. An error of, say, $\pm 10\%$ on these form factors thus reflects in roughly twice the same error on the differential branching ratios. This is illustrated in the left panel of fig. \ref{fig:r_gamma}. Such spread, shown as a blue area, is too large to clearly resolve the $C_{9,10}$ shift required by $R_K$ and $\RKst$. The effect of the latter shift is displayed by the red line in the same figure, and as shown, this line lies barely outside the blue area.

\item Form-factor uncertainties cancel to a large extent in $r_\gamma$. In fact, the $r_\gamma$ variation due to these uncertainties is suppressed by powers of the difference $(m_\mu^2 - m_e^2)/m_{B_s}^2$. The residual theoretical uncertainty amounts to a relative error on $r_\gamma$ of at most 5\%,\footnote{The error depends on the degree of correlation between the form-factor errors. For example, the case of $\FV$ and $\FA$ errors exactly anti-correlated obviously amounts to an additional cancellation -- between the coefficients of the $\FV^2$ and $\FA^2$ terms within each of the two branching ratios in $R_\gamma$. The figure displays the least favorable case, and as such the blue area represents the envelope of any realistic theoretical error on the form factors.} well below the size of the shifts to $C_{9,10}^{(\prime)}$ required by $R_K$ and $\RKst$. This point is illustrated in the right panel of fig. \ref{fig:r_gamma}. In this plot the red line lies well outside the blue band of the theoretical error throughout the considered $q^2$ range.

\item We do not consider $\sh$ values above 0.8. In fact, for such values the FSR component -- diagrams 3 and 4 in fig. \ref{fig:diags} -- becomes, in the $\mumu$ channel, comparable in size with the ISR one -- all the other diagrams in the same figure.\footnote{In the $e^+ e^-$ channel the ISR component stays negligible up to $q^2$ very close to the endpoint, because of chiral suppression. The relative size of the ISR and FSR components can be inferred from fig. 1 of Ref. \cite{Dettori:2016zff}.} Being ISR diagrams proportional to just the first of the matrix elements in eq. (\ref{eq:ffs}), they steadily spoil the cancellation of form-factor uncertainties between the numerator and denominator of eq. (\ref{eq:R_gamma}) as $q^2$ increases. This can be appreciated in fig. \ref{fig:r_gamma} (right panel), where from left to right $r_\gamma$ gradually departs from unity and its error gets larger. Incidentally, this departure from unity distinguishes $\bsllgamma$ decays from $B \to V \ell \ell$ ones.

\end{itemize}

\subsubsection*{Impact of broad charmonium}

\noindent In the discussion so far we have disregarded one further source of potentially significant theory systematics, namely the contamination of the \bsmumugamma spectrum by broad-charmonium resonances. A dedicated study in the context of $B^+ \to K^+ \ell \ell$ has been performed in \cite{Lyon:2014hpa}, and extended by LHCb to include low-lying $\rho,\omega,\dots$ resonances \cite{Aaij:2016cbx}. Similar effects are possible for our decays of interest, through the subprocess $B^0_s \to V_{c \bar c} (\to \ell \ell) \gamma$, with $V_{c \bar c}$ any of $\psi(2S)$, $\psi(3770)$, $\psi(4040)$, $\psi(4160)$ or $\psi(4415)$. We model LD effects associated with such resonances as a sum over Breit-Wigner poles \cite{Kruger:1996cv}, through the replacement
\beq
\label{eq:Vcc_shift}
C_9 ~\to~ C_9 ~-~ \frac{9 \pi}{\alpha^2} \, \bar C \,  \sum_V 
 |\eta_V| e^{i \delta_V} \frac{\hat{m}_V \, \mc B(V \to \mu^+ \mu^-) \, \hat{\Gamma}_{\textrm{tot}}^V}{\hat{q}^2 - \hat{m}_V^2 + i \hat{m}_V \hat{\Gamma}^V_{\textrm{tot}}}~,
\eeq
with free floating absolute value and phase \cite{Lyon:2014hpa} to measure the deviation from naive factorisation ($|\eta_V| = 1$ and $\de_V = 0$).%
\footnote{In  Ref. \cite{Lyon:2014hpa}, for $B \to K \mu^+ \mu^-$ it was found that $|\eta_V| \simeq 2.5$ and $\de_V \simeq \pi$ gives a good description of the data.} The sum runs over the five resonances mentioned above, hats indicate that the given quantity is made dimensionless by an appropriate power of $M_{B_s}$, and $\bar C = C_1 + C_{2}/3 + C_3 + C_{4}/3 + C_5 + C_{6}/3$. The relevant numerical input for all resonances but the $\psi(2S)$ is taken from the recent determination \cite{Ablikim:2007gd}. For the $\psi(2S)$ we use data from the PDG\cite{Olive:2016xmw} and  checked the stability of our results against numerical input taken from \cite{Lyon:2014hpa,Kruger:1996cv}. The effect of the shift (\ref{eq:Vcc_shift}) is shown in either panel of fig. \ref{fig:r_gamma} as a wiggly solid line using $|\eta_V| = 1$ and $\de_V = 0$ for illustrative purposes. The figure displays that, while the differential branching ratio (left panel) shows some sensitivity to such effect, that may partly compensate the new-physics shift required by $R_K$ and $\RKst$, this sensitivity is substantially reduced in $r_\gamma$ (right panel), that neatly distinguishes the SM case from the new-physics one.

\section{Outlook} \label{sec:outlook}

Here we collect considerations on a few further directions that may be promising with future data, and with further thinking.

First, the low-$q^2$ spectrum is in principle sensitive also to $F_{V,A} C_{10}$, as can be appreciated by inspection of the $\bsmumugamma$ differential branching ratio \cite{Melikhov:2004mk}. This sensitivity can be `projected out' by measuring the first $\cos \theta$ moment of the differential distribution, namely the quantity
\beq
\int d q^2 \, d \cos \theta \frac{d \mc B(\bsmumugamma)}{d q^2 \, d \cos \theta} \cos \theta~,
\eeq
where, we remind, $\theta$ is the angle between the three-momentum vectors of the $\mu^-$ and of the photon in the dilepton center-of-mass system \cite{Kruger:2002gf}. By choosing the integration region to be a symmetric interval around the $\phi$ peak, one is able to cancel out the resonant terms and single out the interference ones. Such an analysis is the only way we are aware of to disentangle the different spectrum components -- in particular direct emission versus interference, because terms $\propto \cos^3 \theta$ and higher only enter the latter contributions, and not the former. We also note explicitly that, per definition, interference terms permit access to combined information about {\em all} the Wilson coefficients, and not only the e.m.-dipole ones.

The most outstanding limitation of such a measurement is of course statistical. A crude estimate of the number of events to be expected in the near future for such an analysis can be obtained by projecting the number of signal candidates in Ref. \cite{Aaij:2012ita} to the end of Run 3, and suitably rescaling for the relative branching ratio to the $\mumu$ over the $K^+ K^-$ channels. We thereby estimate around 350 signal candidates for the $\phi \to \mumu$ channel alone. Furthermore, this estimate assumes the same trigger and reconstruction efficiencies as Run 1, which is rather pessimistic.

As concerns the $q^2$ resolution required for such a measurement, it would seem that 
it need to be comparable to, or better than, the width of the resonance in question, which in our $\phi(1020)$ example is about 4 MeV. Such resolution is actually realistic (see e.g. supplementary material in Ref. \cite{Aaij:2016cbx}). Furthermore, even for a resolution larger than the natural width of the resonance, the assignment of events to either side of the peak may be performed {\em statistically} (rather than on an event-by-event basis), because the template for the $q^2$ resolution can be measured elsewhere. Of course the actual effectiveness of such technique requires a dedicated MonteCarlo.\footnote{We thank F. Dettori for a clarifying discussion on this matter.}

An obvious question is whether and to what extent all the above considerations may be applicable in the narrow-charmonium region as well. One first objection is the fact that $\mc B(B_s^0 \to J/\psi \, \gamma)$ is yet to be measured. Furthermore, the presence of the nearby $\psi(2S)$ resonance makes it necessary to also determine the phase of each relevant amplitude.

A further, short comment concerns the $B_d^0$ counterparts to the decays discussed in this paper. They would statistically suffer from the relative CKM suppression, of about $4 \times 10^{-2}$, with respect to the $B^0_s$ modes, {\em but} in principle enormously benefit from the huge statistics and detector capabilities foreseen at Belle 2, although this can be ascertained only through a dedicated study. From the theoretical point of view, many of the considerations we made for the $B_s^0$ modes can be extended to the $B_d^0$ ones. For example the isolated $\phi$ resonance is replaced by the $\rho^0, \omega, ...$ on which, again, data exist. This topic requires separate consideration.

\section{Summary and Conclusions} \label{sec:summary}

In this paper we reappraised the decay $\bsmumugamma$, discussing its main sources of theoretical error as well as strategies towards substantially reducing them. These uncertainties are different in the two kinematical regions that one can realistically exploit at present, and defined by the dilepton invariant mass squared  being below or respectively above the narrow-charmonium region.

In the low-$q^2$ region, the decay is dominated by the subprocess $B_s^0 \to \phi \gamma$, with the $\phi$ decaying to a $\mumu$ pair. Since, however, the process $B_s^0 \to \phi (\to K K) \gamma$ is measured to already good accuracy, we show how the main form-factor uncertainties in this region can be traded for experimental data. We emphasise that our aim in proposing this trade-off is not to better control \bsmumugamma in the region where the amplitude is completely dominated by the $\phi$ resonance. In other words, we are not proposing to use \bsmumugamma as a proxy of $B_s^0 \to \phi \gamma$, because the $\phi \to KK$ decay mode is statistically more advantageous, by far. We rather aim at constraining as well as possible the resonant region with experimental data, in order to better predict the rest of the \bsmumugamma spectrum at low $q^2$. In particular, one may thereby extract the interference terms in the amplitude. We likewise discussed the potential impact of further resonances such as the $\phi^\prime$.

In the high-$q^2$ region, conversely, we exploit the fact that the decay is dominated by two form-factor combinations, plus contributions from broad charmonium that we model accordingly. We construct the ratio $R_\gamma$, akin to $R_K$ and likewise sensitive to lepton-universality violation. We show that in an appropriately chosen $q^2$ range most of the form-factor uncertainties indeed cancel, thereby making $R_\gamma$ sensitive to the Wilson-coefficients' shifts hinted at by $R_K$ and $\RKst$. Our choice of the $q^2$ range takes into account the two main effects we are aware of, that can spoil the cancellation of the form-factor errors in $R_\gamma$: the pollution by broad-charmonium resonances, and the increasing importance of the bremsstrahlung component to the amplitude for high enough $q^2$ in the $\mumu$ channel.

We conclude with predictions of the total branching ratio for \bsmumugamma for low and high $q^2$
\beqa
\label{eq:Bsmumugamma_pred}
&&\mc B(\bar B_s^0 \to \mumu \gamma)_{{\rm low}\,q^2} ~=~ (8.4 \pm 1.3) \times 10^{-9}~,\nn \\
[0.2cm]
&&\mc B(\bar B_s^0 \to \mumu \gamma)_{{\rm high}\,q^2} ~=~ (8.90 \pm 0.98) \times 10^{-10}~,
\eeqa
where the $q^2$-integration windows are respectively $q^2 / M_{B_s}^2 = [(2 m_\mu / M_{B_s})^2, 0.30]$ and $q^2 / M_{B_s}^2 = [0.55, 1 - 2 E_{\rm cut}/M_{B_s}]$, with $E_{\rm cut} = 50$ MeV \cite{Buras:2012ru}.\footnote{The $q^2$ endpoint takes into account that experiments cannot be completely photon-inclusive. The choice we made is indicative. In practice, the best procedure is to extrapolate to the kinematic endpoint by an experimental Monte Carlo, which would also take into account soft resummed bremsstrahlung, cf. footnote~\ref{foot:PHOTOS}.} These ranges correspond to $q^2 = [0.04, 8.64]$ GeV$^2$ and $q^2 = [15.84, 28.27]$ GeV$^2$. For the ratio $R_\gamma$ we obtain
\beq
\label{eq:R_gamma_pred}
R_\gamma(q_1^2, q_2^2) ~=~ 1.152 \pm 0.030~,
\eeq
where, we remind, we choose $q_1^2 / M_{B_s}^2 = 0.55$ and $q_2^2 / M_{B_s}^2 = 0.8$. The errors on the above predictions are obtained by assuming form factors with uncorrelated Gaussian errors of 10\% and $\sqrt (|A^{\bar{B}^0_s \to \phi}_\perp|^2 + |A^{\bar{B}^0_s \to \phi}_\parallel|^2)$ traded for eq. (\ref{eq:Bsphigamma_exp}).

In our above estimates we neglect the possible contribution from the $\phi'$, having estimated it to be below $1\%$. On the other hand, we do include possible systematic effects due to the $J/\psi$ (low $q^2$) or to broad-charmonium resonances (high $q^2$). These effects are modelled according to eq. (\ref{eq:Vcc_shift}), where we take the resonance couplings to be uniformly distributed in the ranges $|\eta_V| \in [1, 3]$ and $\delta_V \in [0, 2 \pi]$, and uncorrelated with one another. The possible pollution from the $J/\psi$ resonance is actually the reason why we limited the low-$q^2$ prediction to $\sh \le 0.30$.\footnote{For reference, taking the low-$q^2$ range to extend up to $\sh \le 0.33$, as chosen elsewhere in the literature, we find
\begin{alignat}{5}
\label{eq:BR0.33}
&\mc B(\bar B_s^0 \to \mumu \gamma)_{\sh \le 0.33}& &~=~& (8.3 \pm 1.3) \times 10^{-9} ~~~~~~~&\mbox{(no $J/\psi$)} \nn \\
&\mc B(\bar B_s^0 \to \mumu \gamma)_{\sh \le 0.33}& &~=~& (10.6 \pm 2.9) \times 10^{-9} ~~~~~~~&\mbox{(with $J/\psi$)}~,
\end{alignat}
where in the first equation we take $|\eta_{J/\psi}| = 0$ in eq. (\ref{eq:Vcc_shift}). Whereas the $J/\psi$ effect is sizeable if the integration range extends up to $\sh = 0.33$, the impact for $\sh \le 0.30$ is well within our quoted uncertainties, as can be inferred from the error bands in eqs. (\ref{eq:BR0.33}) and (\ref{eq:Bsmumugamma_pred}).}

\section*{Acknowledgements}
We are grateful to Gerhard Buchalla, Francesco Dettori and Dmitri Melikhov for important insights. We also acknowledge useful feedback from Olivier Deschamps, Mikolaj Misiak, Francesco Polci and Peter Stangl. DG and RZ also thank Gino Isidori for kind hospitality at the University of Zurich, where part of this work was completed. The work of DG is partially supported by the CNRS grant PICS07229.

\appendix

\section{Status of $B_s^0 \to \ga$ form factors} \label{app:FF}

This appendix aims to summarise the status of the $B_s^0 \to \ga$ form factors $\FVA(q^2)$ and $\FTVA(q^2,k^2)$, defined in eq. \eqref{eq:ffs}, and provide motivation to improve on them in future computations. 

First, we remind the reader that the tensor form factors $\FTVA(q^2,k^2)$ are only needed in either the $\FTVA(q^2,0)$ or $\FTVA(0,q^2)$ configuration if one final-state photon is on-shell. The computation of $\FTVA(0,q^2)$, which actually is a $B_s^0 \to \ga^*$ form factor, is a formidable task since there are cuts in the very low-$q^2$ region due to meson and multi-meson states coupling to the $\bar s s$-state which is best treated in a dispersive approach as outlined in eq.~\eqref{eq:disprel}. 

Let us therefore turn our attention to the form factors $\FVA(q^2)$ and $\FTVA(q^2,0)$. A coherent discussion of these form factors summarising the status in 2002 has been given in  \cite{Kruger:2002gf}. These authors have pointed out the previously mentioned algebraic relation $\FTV(0,0) = \FTA(0,0)$ and built upon the large-energy relations \cite{Charles:1998dr} and various low-$q^2$ evaluations using QCD sum rules and quark models. A single-pole parametrisation has been assumed (reminiscent of vector meson dominance) to be valid in the entire $q^2$ regime. We use \cite{Kruger:2002gf} as our reference parametrisation.

In 2003 the $\FVA(q^2)$ form factors were computed, with the aim to describe $B_u \to \ell \nu \gamma$, in QCD factorisation and at next-to-leading twist (including the photon distribution amplitude) using LCSR by \cite{DescotesGenon:2002mw} and \cite{Ball:2003fq} respectively. In the context of $B \to V$ form factors \cite{Ball:2004rg} it became clear that, unlike for the pion form factor, a single-pole parametrisation is insufficient in $B$ to light 
form factors. In order to counterbalance the possible model dependence of our first parametrisation, we tried a second one using the form-factor results in the $B \to V$ update \cite{Straub:2015ica} as follows. First, one notes that, because of the null photon mass, $\FTVA(q^2,0)$ can be simply related \cite{Dimou:2012un} to the form factors $T_{1,2,3}(q^2)$. Specifically, the zero mass of the photon establishes the relation $T_2(q^2) = T_3(q^2) (1 - q^2 / M^2)$, with $M$ the initial-state meson mass. Explicitly
\begin{alignat}{4}
\label{eq:FTV_BSZ}
& \FV^{\bar{ B}^0_s \to \ga} &\; \leftrightarrow \;& 2 V^{\bar{B}^0_s \to \phi}~, \quad  
& & \FTV^{\bar{ B}^0_s \to \ga} &\;\leftrightarrow \;& 2 T_1^{\bar{B}^0_s \to \phi}~, 
  \nn \\[0.2cm]
& \FA^{\bar{ B}^0_s \to \ga} &\;\leftrightarrow \;& 2 A_1^{\bar{B}^0_s \to \phi}/(1- q^2/M^2_{B_s})~, \quad
& & \FTA^{\bar{B}^0_s \to \ga} &\; \leftrightarrow \;& 2 T_2^{\bar{B}^0_s \to \phi}/(1- q^2/M^2_{B_s}) \;.
\end{alignat}
Second, at the twist-2 level the identification is direct if all higher partial waves of the $\phi$ distribution amplitudes (i.e. Gegenbauer moments other than the decay constants) are neglected, which is a reasonable approximation in the given context. The $q^2$ behaviour of the $V$, $A_1$, $T_1$ and $T_2$ form factors has been studied in Ref. \cite{Straub:2015ica} through a combined LCSR and lattice \cite{Horgan:2013hoa} fit to $B \to K^*$ form factors. Hence use of the data points in \cite{Straub:2015ica} plus eq. (\ref{eq:FTV_BSZ}) would provide a second parameterisation, under the assumption that $B \to K^*$ form-factor data can be used as a proxy for $B \to \gamma$ ones, which as mentioned are to date missing. In support of this assumption, we may quote the fact that, at leading twist, a photon and a $K^*$ couple the same way at high energies.

Nonetheless we find substantial differences between the two above-mentioned form-factor sets at high $q^2$. This matter can be settled only by a first-principle computation of $B^0_s \to \gamma$ form factors, that to the best of our knowledge is still missing, and as such highly desirable.

In summary, considerable work is still to be done for these form factors, at low and at 
high $q^2$ alike. The former gap can be filled by LCSR and the latter by lattice QCD.

\bibliographystyle{JHEP}
\bibliography{note_Bsmumugamma_direct}

\end{document}